\documentclass[11pt,a4paper,oneside]{article}
\usepackage[utf8]{inputenc}
\usepackage{amsmath}
\usepackage{amsfonts}
\usepackage{amssymb}
\usepackage{indentfirst}
\usepackage{graphicx}
\usepackage[outdir=figures/]{epstopdf}

\graphicspath{{figures/}}
\usepackage{authblk}
\usepackage{multirow}

\title{An implicit $ P $-multigrid flux reconstruction method for simulation of locally preconditioned unsteady Navier--Stokes equations at low Mach numbers}
\author{Lai Wang\thanks{PhD candidate. Email: bx58858@umbc.edu}}
\author{Meilin Yu\thanks{Assistant Professor. Corresponding author. Email: mlyu@umbc.edu}} 

\affil{Department of Mechanical Engineering\\
	University of Maryland, Baltimore County, Baltimore, MD 21250}

\date{\vspace{-10ex}}
\begin{document}
\maketitle
\section*{Abstract}
We develop a $ P $-multigrid solver to simulate locally preconditioned unsteady compressible Navier--Stokes equations at low Mach numbers with implicit high-order methods. Specifically, the high-order flux reconstruction/correction procedure via reconstruction (FR/CPR) method is employed for spatial discretization and the high-order time integration is conducted by means of the explicit first stage, singly diagonally implicit Runge-Kutta (ESDIRK) method. Local preconditioning is used to alleviate the stiffness of the compressible Navier--Stokes equations at low Mach numbers and is only conducted in pseudo transient continuation to ensure the high-order accuracy of ESDIRK methods.  We employ the element Jacobi smoother to update the solutions at different $ P $-levels in the $ P $-multigrid solver.
High-order spatiotemporal accuracy of the new solver for low-Mach-number flow simulation is verified with the isentropic vortex propagation when the Mach (Ma) number of the free stream is 0.005.
The impact of the hierarchy of polynomial degrees on the convergence speed of the $ P $-multigrid method is studied via several numerical experiments, including two dimensional (2D) inviscid and viscous flows over a NACA0012 airfoil at $\text{Ma} = 0.001$, and a three dimensional (3D) inviscid flow over a sphere at $\text{Ma} = 0.001$.
The $ P $-multigrid solver is then applied to coarse resolution simulation of the transitional flows over an SD7003 wing at $ 8^\circ $  angle of attack when the Reynolds number is 60000 and the Mach number is 0.1 or 0.01.

\section*{Key Words}
$ P $-multigrid; high-order flux reconstruction; implicit time marching; low Mach number; unsteady flows; coarse resolution simulation 

\section{Introduction}
Many industrial applications of fluids concern low speed flows and need a fast turnaround time to obtain reasonable analysis results, especially for turbulent flows. Efficient under-resolved turbulence simulation of low speed flows is a promising approach to meet those industrial requirements.  Due to the superior numerical properties of high-order methods over their low-order counterparts for coarsely resolved computations, many researches~\cite{Gassner2013,Sherwin2015,Bassi2016,Wang2017CF} on under-resolved simulation of turbulent flows have been conducted in recent years. Among them few have discussed the simulation of low-Mach-number flows by solving the compressible Navier--Stokes equations directly, an important approach to tackle flow simulation at all speeds
and to take full advantage of contemporary technique development in computational fluid dynamics (CFD). 
In this study, we aim to develop an efficient high-order numerical framework for coarse resolution simulation of locally preconditioned unsteady compressible Navier-Stokes equations at low Mach numbers.          

There are several challenges needed to be addressed towards achieving our research goal. First, when solving convection-dominated problems at low Mach numbers, the accuracy of numerical methods degrades due to the large disparities between the propagation speeds of different characteristics. Moreover, the nonlinear systems, resulted from the high-order spatiotemporal discretizations of the compressible Navier--Stokes equations, become hard to solve due to the rapidly increasing stiffness as the Mach number decreases to the incompressible range. To address these issues, we employ and synergize local preconditioning techniques~\cite{Turkel1987, TurkelLeer1993, WeissSmith1995, wang2019implicit} for unsteady compressible Navier-Stokes equations, high-order FR/CPR spatial discretization methods~\cite{Huynh2007,Huynh2009,Wang2009,Vincent2011}, high-order implicit time integration methods (i.e., ESDIRK)~\cite{bijl2002implicit,Kennedy2016,bassi2015linearly,wang2019comparative}, and $ P $-multigrid~\cite{bassi2003numerical,fidkowski2005p,nastase2006high,luo2006p,liang2009p}  in this study. We briefly review the state-of-the-art developments of these methods/techniques, and discuss our contribution.  

For low-Mach-number flows, local preconditioning essentially balances the propagation speeds of the characteristics originated from the hyperbolic part of the Navier--Stokes equations. Hence, the convergence of iterative methods can be accelerated and the accuracy of numerical methods can be preserved for low Mach flows as well. We note that in the high-order method community, the artificial compressibility method~\cite{bassi2015linearly,yu2016high,franciolini2017efficiency,loppi2018high} has been widely used for low speed flow simulation under the assumption that the flow is incompressible. In this study, we pursue the local preconditioning method for compressible flows due to its flexibility on solving flows of all speeds. A comprehensive comparison of the artificial compressibility method and local preconditioning method in the context of high-order methods is yet to be done. Some preliminary work can be found in Ref.~\cite{wang2019implicit}.

The FR/CPR method adopted in this study was first developed by Huynh~\cite{Huynh2007,Huynh2009}, and a family of the FR/CPR methods have been substantially developed by many researchers~\cite{Wang2009,Vincent2011,Romero2016,wang2017compact,Huynh2019}. The idea of reconstructing the local solution polynomials by the correction procedure enables the FR/CPR method to recover many other popular high-order methods, such as discontinuous Galerkin (DG)~\cite{cockburn1989tvb,bassi1997high,vanLeer:2005,wang2007implicit,NDG08}, spectral volume (SV)~\cite{Wang2002SV}, and spectral difference (SD)~\cite{Kopriva1996SD,liu2006spectral}.
Recent researches~\cite{HuynhWangVincent2014CF,Wang2017CF,loppi2018high} have demonstrated that high-order FR/CPR methods are promising spatial discretization methods for simulating complex vortex-dominated flows, aeroacoustics and turbulent flows. 

High-order explicit Runge-Kutta methods ~\cite{cockburn1989tvb,gottlieb2001strong} have been widely used with high-order spatial discretizations to achieve high-order spatiotemporal accuracy due to their ease of implementation. Implicit Runge-Kutta methods have also attracted much research interest due to their stability advantages over the explicit methods~\cite{bijl2002implicit,Kennedy2016,bassi2015linearly,wang2019comparative}. When implicit time integrators are employed, Newton-Krylov methods are usually used to solve the large nonlinear/linear systems~\cite{knoll2004jacobian,persson2008newton}. Although varying degrees of success have been achieved, matrix-based Newton-Krylov methods suffer from large memory consumption. By introducing an approximation of the matrix-vector production in Krylov subspace methods, a matrix-free implementation can significantly reduce the memory usage. Matrix-free implementations of Newton-Krylov methods~\cite{knoll2004jacobian} for high-order methods have been extensively studied~\cite{luo2001computation, crivellini2011implicit, birken2012efficient, birken2013preconditioning,sarshar2017numerical,franciolini2017efficiency}. In our recent work~\cite{wang2019comparative}, a comparative study of various ESDIRK, Rosenbrock and backward differentiation formula (BDF) methods has been conducted in the context of matrix-free implementation of Newton-Krylov methods when FR/CPR is employed for the spatial discretization. In this study, we employ ESDIRK methods to carry out high-order time integration.
  
The multigrid method can be a competitive alternative of the Newton-Krylov methods as it can significantly accelerate the convergence speed of classic iterative methods. Jameson~\cite{jameson1983solution} has pioneered in applying the multigrid method to the fast solution of Euler equations, and since then numerous advances have been  made~\cite{mavriplis1990multigrid,peraire1993multigrid,knoll1999multigrid}. An attribute of the high-order methods is that the built-in compact nature enables a straightforward $ P $-multigrid implementation to accelerate the convergence speed of classic iterative methods. The $ P $-multigrid approach has been successfully applied to solve both Euler and Navier--Stokes equations with high-order spatial discretizations in recent decades~\cite{bassi2003numerical,fidkowski2005p,nastase2006high,luo2006p,liang2009p}. Helenbrook et al.~\cite{helenbrook2003analysis} have analyzed the performance of multigrid solvers for both diffusion and convection problems. They found that the anisotropic nature of convection problems can hinder the performance of the isotropic $ P $-multigrid method with an element Jacobi smoother. Fidkowski et al.~\cite{fidkowski2005p} proposed to use the element line Jacobi smoother to improve the convergence of $ P $-multigrid methods for high Reynolds number flow simulation with stretched grids. 
To further accelerate convergence, $ P $-multigrid methods can be combined with the geometric multigrid methods~\cite{nastase2006high,wallraff2015multigrid}. We also note that $ P $-multigrid methods can serve as preconditioners for Newton-Krylov methods~\cite{persson2008newton,franciolini2018p}.

\textit{Contributions}. In this paper, we aim to develop implicit high-order flux reconstruction methods to solve the unsteady compressible Navier--Stokes equations at low Mach numbers with the  $ P $-multigrid acceleration technique. To better understand numerical properties of the $ P $-multigrid solver, numerical experiments have been conducted to study the impact of the polynomial degree hierarchy on the convergence speed of the multigrid solver. This has seldom been conducted in previous works; some preliminary results have been reported in the conference paper~\cite{Wang_Yu2019_pM}. 
Our numerical experiments consistently suggest that a polynomial degree hierarchy close to $ \{P_0-P_0/2-P_0 \} $ or $ \{P_0-P_0/2-P_0/4-P_0/2-P_0\} $ should be employed for the two-level or three-level V-cycle $ P $-multigrid solver to achieve the best convergence acceleration, where $ P_0 $ is the maximum polynomial degree used in the  $ P $-multigrid. The $ P $-multigrid solver has been applied to coarse resolution simulation of the transitional flows over an SD7003 wing at $ \text{Ma}=0.1 $ and $ 0.01 $. 

\textit{Article Organization}. The remainder of the paper is organized as follows. In Section~\ref{GENM}, we first review the local preconditioning method, and then briefly introduce FR/CPR and ESDIRK methods. In Section~\ref{p_MultiGrid}, we explain the $ P $-multigrid method. We then present and discuss numerical results from several 2D and 3D low-Mach-number flow simulations in Section~\ref{Results}. The last section summarizes this work.

\section{Numerical methods} \label{GENM}
\subsection{Governing equations} 
On using Einstein summation convention, the compressible  Navier--Stokes equations can be written as 

\begin{equation}\label{mass}
\frac{\partial \rho}{\partial t} + \frac{\partial( \rho u_j)}{\partial x_j} = 0,
\end{equation}
\begin{equation}\label{momentum}
\frac{\partial (\rho u_i)}{\partial t} + \frac{\partial (\rho u_j u_i+\delta_{ji}p)}{\partial x_j} = \frac{\partial \tau_{ji}}{\partial x_j},
\end{equation}
\begin{equation}\label{energy}
\frac{\partial(\rho E) }{\partial t}+\frac{\partial(\rho  u_j H)}{\partial x_j} = \frac{\partial(u_i\tau_{ij}-K_j)}{\partial x_j},
\end{equation}
where $i= 1 ,\dots ,d$, and  $ d $ is the size of the problem dimension. Herein, $ \rho $ is the fluid density, $ u_i $ is the velocity component, $ p $ is the pressure, $  E=\frac{p/\rho}{\gamma-1}+\frac{1}{2} u_k u_k $ is the specific total energy, $ H=E+\frac{p}{\rho} $ is the specific total enthalpy, $ \tau_{ij} $ is the viscous stress, $ K_j $ is the heat flux, and $\delta_{ij}$ is the Kronecker delta. We note that in the definition of the specific total energy, $\gamma$ is the specific heat ratio defined as $ \gamma = C_{p}/C_{v} $, where $ C_{p} $ and $ C_{v} $ are  specific heat capacity at constant pressure and volume, respectively. In this study, $\gamma$ is set as 1.4. The ideal gas law $ p=\rho RT $ holds, where $ R $ is the ideal gas constant and $ T $ is the temperature. The viscous stress tensor and heat flux vector are given by
\begin{equation}
\tau_{ij} = 2 \mu \left\{S_{ij}-\frac{1}{3}\frac{\partial u_k}{\partial x_k}\delta_{ij}\right\},
\end{equation}
\begin{equation}
K_j = -\frac{\mu C_p}{Pr}\frac{\partial T}{\partial x_j},
\end{equation}
where $ \mu  $ is the fluid dynamic viscosity, Pr is the molecular Prandtl number, and the strain-rate tensor $ S_{ij} $ is defined as 

\begin{equation}
S_{ij} = \frac{1}{2}\left(\frac{\partial u_i}{\partial x_j}+\frac{\partial u_j}{\partial x_i}\right).
\end{equation}	
In this study, $ \mu $ is treated as a constant and Pr is set as $ 0.72 $.

\subsection{Local preconditioning for steady problems}
Eqs.~\eqref{mass},~\eqref{momentum}, and~\eqref{energy} can be rewritten in the symbolic format as
\begin{equation} \label{ns_equation}
\frac{\partial \boldsymbol{q}}{\partial t} + \nabla \cdot \boldsymbol{f} = 0,
\end{equation}
where $ \boldsymbol{q} = (\rho, \rho u_i, \rho E)^T $ are the conservative variables and $ \boldsymbol{f} $ is the flux tensor.
The local preconditioning approach employed in this study uses primitive variables $ (p,u_i,T)^T $ as the working variables~\cite{WeissSmith1995, wang2019implicit}. To avoid confusion, we use $ \boldsymbol{q}_c $ and $ \boldsymbol{q}_p $ to denote the conservative variables $ (\rho,\rho u_i, \rho E)^T $ and primitive variables $ (p,u_i,T)^T $, respectively.
One can apply the chain rule to the temporal derivative in Eq.~\eqref{ns_equation} to obtain 
\begin{equation} \label{ns_equation_pseudo_time_chain_rule}
\boldsymbol{M}\frac{\partial \boldsymbol{q}_p}{\partial t} + \nabla \cdot \boldsymbol{f} = 0,
\end{equation}
where $ \boldsymbol{M} = {\partial \boldsymbol{q}_c}/{\partial \boldsymbol{q}_p} $. Then the Jacobian matrix $ \boldsymbol{M} $ is replaced with the preconditioning matrix $ \boldsymbol{\varGamma }$. For a 3D problem, $ \boldsymbol{\varGamma }$ reads
	\begin{equation}\label{preconditioning}
	\boldsymbol{\varGamma} =
	\begin{pmatrix}
	\Theta    & 0      & 0      & 0      & \rho_{T}   \\
	\Theta u  & \rho   & 0      & 0      & \rho_{T} u \\
	\Theta v  & 0      & \rho   & 0      & \rho_{T} v \\
	\Theta w  & 0      & 0      & \rho   & \rho_{T} w \\
	\Theta H-1& \rho u & \rho v & \rho w & \rho_{T} H + \rho C_{p} 
	\end{pmatrix}.			 
	\end{equation} 
where 
\begin{equation}\label{theta}
\varTheta = \Bigl(\frac{1}{U^{2}_{r}}-\frac{\rho_{T}}{\rho C_{p}}\Bigr).
\end{equation}
Herein, $ U_r $ is the reference velocity, which can be modeled as 
 $ U_r=\epsilon c $, where $c$ is the speed of sound. The free parameter $ \epsilon $ is defined as
\begin{equation}\label{eps_nonmove}
\epsilon = \min(1,\max(\kappa Ma_\infty, Ma)),
\end{equation} 
where $ \kappa$ is a free parameter. The global cut-off parameter  $\kappa Ma_\infty $ is employed to prevent robustness deterioration instabilities near stagnation points. If not specifically mentioned $ \kappa = 1.0 $ for all numerical simulations.
The eigenvalues of the inviscid part of the preconditioned Navier--Stokes equations 
\begin{equation} \label{ns_equation_pseudo_time_preconditioned}
\boldsymbol{\varGamma}\frac{\partial \boldsymbol{q}_p}{\partial \tau} + \nabla \cdot \boldsymbol{f} = 0,
\end{equation}
in the face normal direction $ \boldsymbol{n} $ are $ u_n,u_n,u_n,u_n'+c',u_n'-c' $ where~\cite{WeissSmith1995}

\begin{equation}\label{eigenvalues_para}
\begin{cases}
u_n=\boldsymbol{v}\cdot\boldsymbol{n}\\
u_n'=u_n(1-\alpha)\\
c'=\sqrt{\alpha^{2}u_n^{2}+U_{r}^{2}} \\
\alpha = (1-\beta U_{r}^{2})/2\\
\beta = \left(\rho_{p}+\frac{\rho_{T}}{\rho C_{p}}\right).
\end{cases}
\end{equation}	
Herein, $\boldsymbol{v}$ is the velocity vector. For an ideal gas, $ \beta=(\gamma RT)^{-1}=1/c^{2} $. At low speed, when $ U_{r} $ approaches zero, $ \alpha $ will approach $ \frac{1}{2} $. All the eigenvalues will then have the same magnitude as $ u_n  $. Thus, the stiffness of the compressible Navier--Stokes equations is significantly decreased. Note that the local preconditioning method will destroy the time accuracy of Eq.~\eqref{ns_equation_pseudo_time_chain_rule}. Thus, the pseudo time $ \tau $ is introduced here which is conventionally used in the pseudo transient continuation to solve the nonlinear equations in both steady and unsteady problems.
For more information, the readers are referred to Refs.~\cite{WeissSmith1995, wang2019implicit}.

\subsection{The FR/CPR method with local preconditioning}
For completeness, a brief review of the FR/CPR method~\cite{Wang2009} is presented in this section when local preconditioning is employed.
The preconditioned Navier--Stokes equations~\eqref{ns_equation_pseudo_time_preconditioned}
\begin{equation*} \label{ns_equation_pseudo_time_preconditioned_2}
\boldsymbol{\varGamma}\frac{\partial \boldsymbol{q}_p}{\partial \tau} + \nabla \cdot \boldsymbol{f} = 0,
\end{equation*}
is defined in domain $ \Omega $ which is partitioned into $ N $ non-overlapping elements $ \Omega_e $, where $ e=1,2,\ldots,N  $. 
After multiplying each side by the test function $ \boldsymbol{\vartheta} $ and integrating over $ \Omega_e$, one obtains
\begin{equation} \label{eqn:NS_weighted}
\int_{\Omega_e}\boldsymbol{\varGamma}\frac{\partial  \boldsymbol{q}_{p,e}}{\partial \tau} \boldsymbol{\vartheta }dV + \int_{\Omega_e}\boldsymbol{\vartheta}\nabla \cdot \boldsymbol{f}_edV =0
\end{equation}
On applying the integration by parts and divergence theorem, Eq.~\eqref{eqn:NS_weighted} reads

\begin{equation} \label{eqn:IBP_1}
\int_{\Omega_e}\boldsymbol{\varGamma}\frac{\partial \boldsymbol{q}_{p,e}}{\partial \tau} \boldsymbol{\vartheta }dV + \int_{\partial \Omega_e} \boldsymbol{\vartheta}\boldsymbol{f}_e\cdot \boldsymbol{n} dS -\int_{\Omega_k}\boldsymbol{f}_e\cdot\nabla  \boldsymbol{\vartheta }dV=0,
\end{equation}
where $ \boldsymbol{n} $ is the outward-going normal direction of the faces of the element $ \Omega_e $. In the discrete form, we assume $ \boldsymbol{q}_{p,e}^h $ is the approximate solution in element $ \Omega_e $. The solution and the test function belong to the polynomial space of degree $ k $, i.e., $  \boldsymbol{q}_{p,e}^h \in P^k $ and $  \boldsymbol{\vartheta}^h \in P^k $. The ensure conservation, $ \boldsymbol{f}_e\cdot \boldsymbol{n} $ in Eq.~\eqref{eqn:IBP_1} is replaced with $ \boldsymbol{f}^{com}_{\boldsymbol{n}} $, the common flux in the normal direction of the element surfaces. Eq.~\eqref{eqn:IBP_1} then reads 
\begin{equation}\label{numerical_ibp}
\int_{\Omega_e}\boldsymbol{\varGamma}\frac{\partial \boldsymbol{q}_{p,e}^h}{\partial \tau} \boldsymbol{\vartheta }^hdV + \int_{\partial \Omega_e} \boldsymbol{\vartheta}^h\boldsymbol{f}^{com}_{\boldsymbol{n}} dS -\int_{\Omega_k}\boldsymbol{f}_e^h\cdot\nabla  \boldsymbol{\vartheta }^hdV=0.
\end{equation}
After applying integration by parts and divergence theorem again to the last term of Eq.~\eqref{numerical_ibp}, one obtains
\begin{equation}\label{penalty}
\int_{\Omega_e}\boldsymbol{\varGamma}\frac{\partial \boldsymbol{q}_{p,e}^h}{\partial \tau} \boldsymbol{\vartheta}^h dV + \int_{\Omega_e}\boldsymbol{\vartheta}^h\nabla \cdot \boldsymbol{f}_e^hdV+ \int_{\partial \Omega_e} \boldsymbol{\vartheta}^h[ \boldsymbol{f} ] dS=0,
\end{equation}
where $ [\boldsymbol{f}]=\boldsymbol{f}^{com}_{\boldsymbol{n}} - \boldsymbol{f}^{loc}_{\boldsymbol{n}} $ with $ \boldsymbol{f}^{loc}_{\boldsymbol{n}} = \boldsymbol{f}_e^h\cdot \boldsymbol{n}$.
In FR/CPR, the correction field $ \boldsymbol{\delta}_e \in P^k $ is defined as~\cite{Wang2009}
\begin{equation}
\int_{\partial \Omega_e} \boldsymbol{\vartheta }^h[\boldsymbol{f} ] dS = \int_{\Omega_e} \boldsymbol{\vartheta }^h \boldsymbol{\delta}_edV.
\end{equation}
Therefore, Eq.~\eqref{penalty} can be expressed as
\begin{equation}
\int_{\Omega_e}\left(\boldsymbol{\varGamma}\frac{\partial \boldsymbol{q}_{p,e}^h}{\partial \tau} + \nabla\cdot\boldsymbol{f}_e^h+\boldsymbol{\delta}_e\right) \boldsymbol{\vartheta}^h dV =0.
\end{equation}
The differential form can then be obtained as
\begin{equation} \label{eqn:FR_diff}
\boldsymbol{\varGamma}\frac{\partial \boldsymbol{q}_{p,e}^h}{\partial \tau} + \mathbb{P}\left(\nabla\cdot\boldsymbol{f}_e^h\right)+\boldsymbol{\delta}_e =0.
\end{equation}
Herein,  $ \mathbb{P} \left(\nabla\cdot\boldsymbol{f}_e^h\right)$ is the projection of the flux divergence $ \left(\nabla\cdot\boldsymbol{f}_e^h\right)$, which may not be a polynomial, onto an appropriate polynomial space. We note that Eq.\eqref{eqn:FR_diff} can be directly derived from the differential form of the governing equations; their equivalence has been established in Ref.~\cite{Yu_Wang_2013}.
Specifically, for quadrilateral and hexahedral elements, the correction field can be obtained by means of tensor product of the one dimensional correction polynomials; for triangular and tetrahedral elements, the readers are referred to Ref.~\cite{WilliamsEtal2013JCP}. Only quadrilateral and hexahedral elements are considered in this study. 

A key step to solve Eq.~\eqref{eqn:FR_diff} is to construct the common normal flux $ \boldsymbol{f}^{com}_{\boldsymbol{n}} $. The approximate Riemann solver in Ref.~\cite{WeissSmith1995} is used to calculate the common inviscid fluxes at the element interfaces in their normal directions as
\begin{equation}\label{invis_com}
\boldsymbol{f}_{\boldsymbol{n},inv}^{com} = \frac{\boldsymbol{f}_{\boldsymbol{n},inv}^{+}+\boldsymbol{f}_{\boldsymbol{n},inv}^{-}}{2}-\boldsymbol{\varGamma}\boldsymbol{R}|\boldsymbol{\varLambda}|\boldsymbol{R}^{-1}\frac{\boldsymbol{q}^{+}_p-\boldsymbol{q}^{-}_p}{2},
\end{equation}
where superscripts `$-$' and `$+$' denote the left of right side of the current interface, the subscript $ \boldsymbol{n} $ is the unit normal direction from left to right, $\boldsymbol{\varLambda}$ is a diagonal matrix consisting of the eigenvalues of the preconditioned Jacobian $\boldsymbol{\varGamma}^{-1}\partial \boldsymbol{f}_{\boldsymbol{n}} /  \partial \boldsymbol{q}_p$, and $\boldsymbol{R}$ consists of the corresponding right eigenvectors evaluated with the averaged values. 
The common viscous fluxes at the element interfaces are $ \boldsymbol{f}_{n,vis}^{com} = \boldsymbol{f}_{vis}(\boldsymbol{q}^{+}_p,\nabla \boldsymbol{q}^{+}_p,\boldsymbol{q}^{-}_p,\nabla \boldsymbol{q}^{-}_p) $. Here we need to define the common solution $ \boldsymbol{q}^{com}_p $ and common gradient $ \nabla \boldsymbol{q}^{com}_p $ at the cell interface. 
On simply taking average of the primitive variables, we get 
\begin{equation}\label{com_q}
\boldsymbol{q}^{com}_p = \frac{\boldsymbol{q}^{+}_p+\boldsymbol{q}^{-}_p}{2}.
\end{equation}
The common gradient is computed as 
\begin{equation}\label{com_grad_q}
\nabla \boldsymbol{q}^{com}_p = \frac{\nabla\boldsymbol{q}^{+}_p+\boldsymbol{r}^{+}_p+\nabla\boldsymbol{q}^{-}_p+\boldsymbol{r}^{-}_p}{2},
\end{equation}
where $ \boldsymbol{r}^{+} $ and $ \boldsymbol{r}^{-} $ are the corrections to the gradients on the interface. The second
approach of Bassi and Rebay (BR2)~\cite{Bassi2005} is used to calculate the corrections. 

\subsection{ESDIRK with dual-time stepping}

The ESDIRK methods for the 3D compressible Navier-Stokes equation~\eqref{ns_equation} can be written as 
\begin{equation}\label{ESDIRK_ns}
\begin{cases}
\boldsymbol{q}_c^{n+1} = \boldsymbol{q}_c^{n}+\Delta t
\sum_{i=1}^{s}b_i\boldsymbol{R}(\boldsymbol{q}_c^i),\\
\boldsymbol{q}_c^i = \boldsymbol{q}_c^n, i=1,\\
\boldsymbol{q}_c^i = \Delta t \omega \boldsymbol{R}(\boldsymbol{q}_c^i)+\boldsymbol{q}_c^n+\Delta t \sum_{j=1}^{i-1}a_{ij}\boldsymbol{R}(\boldsymbol{q}_c^j),i=2,\dots,s,
\end{cases}	
\end{equation}
where $ i$ is the stage number, $ s $ is number of total stages, $ n $ denotes the physical time step and $ \boldsymbol{R} = -\nabla\cdot\boldsymbol{f} $. The second-order, three-stage ESDIRK2~\cite{Kennedy2016}, third-order, four-stage ESDIRK3~\cite{bijl2002implicit} and fourth-order, six-stage ESDIRK4~\cite{bijl2002implicit} methods are studied in this paper. Note that the temporal discretization is for conservative variables. In every stage except the first one, a nonlinear system is to be solved, which can be expressed as 	
\begin{equation}\label{ESDIRK_conservative}
\boldsymbol{F}(\boldsymbol{q}_c^{i}) = \left(-\frac{1}{\omega\Delta t}\boldsymbol{q}_c^i +\boldsymbol{R}(\boldsymbol{q}_c^i)\right)+\frac{1}{\omega\Delta t}\left(\boldsymbol{q}_c^n+\Delta t\sum_{j=1}^{i-1}a_{ij}\boldsymbol{R}(\boldsymbol{q}_c^j)\right),i=2,\dots,s.
\end{equation}
Since $ \boldsymbol{q}_c $ is de facto a function of $ \boldsymbol{q}_p $, Eq.~\eqref{ESDIRK_conservative} can be reformulated as 
\begin{equation}\label{ESDIRK_primitive}
\boldsymbol{F}(\boldsymbol{q}_{p}^i) = \left(-\frac{1}{\omega\Delta t}\boldsymbol{q}_{c}^i(\boldsymbol{q}_p) +\boldsymbol{R}(\boldsymbol{q}_{p}^i)\right)+\frac{1}{\omega\Delta t}\left(\boldsymbol{q}_c^n(\boldsymbol{q}_p)+\Delta t\sum_{j=1}^{i-1}a_{ij}\boldsymbol{R}(\boldsymbol{q}^j_p)\right),\ i=2,\dots,s.
\end{equation}
A dual-time stepping procedure for the $ i $-th stage reads

\begin{equation}\label{pseudo_transient}
\boldsymbol{\varGamma}\frac{\boldsymbol{q}^{m+1,i}_p-\boldsymbol{q}^{m,i}_p}{\Delta \tau}=\boldsymbol{F}(\boldsymbol{q}^{m+1,i}_p),
\end{equation}
where $ m $ is the iteration step for the pseudo-transient continuation. This procedure can ensure that the preconditioning is only enforced in the pseudo-time marching. Therefore, the accuracy of ESDIRK  can be preserved.
Eq.~\eqref{pseudo_transient} can be linearized as 
\begin{equation}\label{linearized_pseudo_transient_final}
\left(\frac{\boldsymbol{\varGamma}}{\Delta \tau} +\frac{\boldsymbol{M}}{\omega \Delta t}-\frac{\partial \boldsymbol{R}}{\partial\boldsymbol{q}_p}\right)^m\Delta \boldsymbol{q}^{m,i}_p= \boldsymbol{F}(\boldsymbol{q}^{m,i}_p),
\end{equation}
where $ \Delta \boldsymbol{q}^{m,i}_p = \boldsymbol{q}^{m+1,i}_p-\boldsymbol{q}^{m,i}_p $.
As a result, the solution can be updated as
\begin{equation}\label{smoothing}
\boldsymbol{q}_p^{m+1,i} = \boldsymbol{q}_p^{m,i} + \boldsymbol{A}^{-1}\boldsymbol{F}(\boldsymbol{q}_p^{m,i}),
\end{equation}
where $ \boldsymbol{A} =  \left(\frac{\boldsymbol{\varGamma}}{\Delta \tau} +\frac{\boldsymbol{M}}{\omega \Delta t}-\frac{\partial \boldsymbol{R}}{\partial\boldsymbol{q}_p}\right)^m$.
In order to save the memory usage, the smoothing step is conducted as 
\begin{equation}\label{smoothing_element_jacobi}
\boldsymbol{q}_p^{m+1,i} = \boldsymbol{q}_p^{m,i} +\alpha_r \boldsymbol{D}^{-1}\boldsymbol{F}(\boldsymbol{q}_p^{m,i}),
\end{equation}
where $ \boldsymbol{D} $ is the block diagonal matrix of $ \boldsymbol{A} $ and $ \alpha_r $ is a relaxation parameter which is set as one  in this study. The smoother is referred to as the element Jacobi smoother. We employ a modified successive evolution relaxation (SER) algorithm~\cite{mulder1985experiments} to update the Courant-Friedrichs-Lewy (CFL) number as  
\begin{equation}\label{ser}
CFL^{0} = CFL_{init},
CFL^{m+1} = \min\left(CFL^{m}\left(\frac{||\boldsymbol{F}||_{L_2}^{m-1}}{||\boldsymbol{F}||_{L_2}^{m}}\right)^{1.5}, CFL_{max} \right).
\end{equation}
The CFL number in this study is calculated from 
\begin{equation}
CFL = \frac{\Delta \tau}{\Delta \tau_{min}},
\end{equation}
where $ \Delta \tau_{min} $ is $ \Delta \tau_{min} = \min(\Delta \tau_{inv}, \Delta \tau_{vis}) $ with $ \Delta \tau_{inv} $ and $ \Delta \tau_{vis} $ defined as
\begin{equation}
\Delta \tau_{inv} = \min\left\{\left(\frac{\Delta x/(P+1)}{c+|\boldsymbol{v}|}\right)_e\right\},\ \text{and} \ \Delta \tau_{vis} = \min\left\{\left(\frac{[\Delta x/(P+1)]^2}{\mu/\rho}\right)_e\right\}.
\end{equation}
$ \Delta x $ is twice as the minimum distance of the barycenter of element $ \Omega_e $ to its surfaces. 
We note that the growth ratio in Eq.~\eqref{ser} is fixed at 1.5; a larger value will make the SER method more aggressive. In order to avoid instabilities, we only employ moderately large $ CFL_{max} $ or $ \Delta \tau_{max} $ in this study.

\section{The $ P $-multigrid method}\label{p_MultiGrid}
To avoid confusion, we neglect the subscript `$ p $' in the working variable $ \boldsymbol{q}_p $ in this section.
Consider a typical three level V-cycle $ P $-multigrid method. The hierarchy of the polynomial degrees is $ \{P_0-P_1-P_2-P_1-P_0\} $, where $ P_0 $ is the maximum polynomial degree used in the cycle. One needs to solve a nonlinear system at each level expressed as 
\begin{equation}\label{lvl_p_0}
\boldsymbol{F}_{P_0}(\boldsymbol{q}_{P_0})-\boldsymbol{S}_{P_0}=0,
\end{equation}
\begin{equation}\label{lvl_p_1}
\boldsymbol{F}_{P_1}(\boldsymbol{q}_{P_1})-\boldsymbol{S}_{P_1}=0,
\end{equation}
\begin{equation}\label{lvl_p_2}
\boldsymbol{F}_{P_2}(\boldsymbol{q}_{P_2})-\boldsymbol{S}_{P_2}=0,
\end{equation}
where $ \boldsymbol{F} $ is defined in Eq.~\eqref{ESDIRK_primitive} for unsteady problems and  $ \boldsymbol{F} = -\nabla \cdot \boldsymbol{f} $ for steady problems. The subscripts $ P_0 $, $ P_1 $ and $ P_2 $ denote the polynomial degrees at the corresponding level. $ \boldsymbol{S}_{P_0} $, $ \boldsymbol{S}_{P_1} $ and $ \boldsymbol{S}_{P_2} $ are referred to as  forcing terms. 
Note that \begin{equation}
\boldsymbol{S}_{P_0} = 0.
\end{equation} 
The procedure of a three-level V-cycle $ P $-multigrid method is illustrated in Figure~\ref{vcycle_fig}. The superscript `b' means before smoothing, `a' means after smoothing and `c' means corrected solution at the current $ P $-level. Specifically, the procedure of a typical three-level V-cycle $ P $-multigrid method can be organized as follows~\cite{fidkowski2005p,luo2006p,liang2009p}:
\begin{figure}		
	\centering
	\includegraphics[width=10cm]{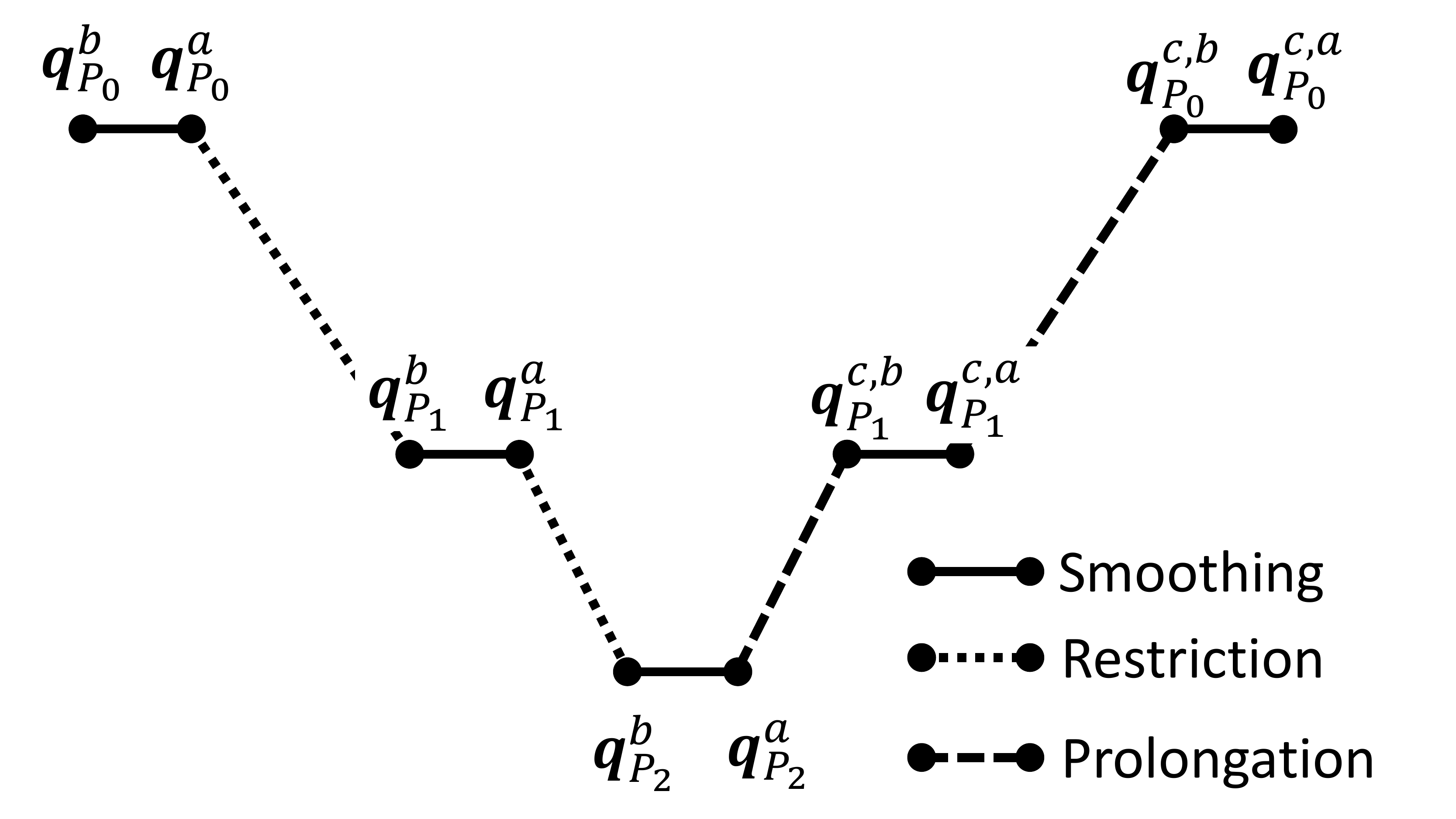} 
	\caption{Illustration of a typical three-level V-cycle of the $ P $-multigrid method.}	
	\label{vcycle_fig}	
\end{figure}

\begin{itemize}
\item
Before smoothing, the initial value of $ \boldsymbol{q} $ at the first level is $ \boldsymbol{q}_{P_0}^b $.
Smooth Eq.~\eqref{lvl_p_0} using the element Jacobi smoother Eq.~\eqref{smoothing_element_jacobi} for a few steps.  The primitive variables  after smoothing is expressed as $ \boldsymbol{q}^{a}_{P_0}$.

\item The defect at the first level 
\begin{equation}
 \boldsymbol{d}_{P_0} = \boldsymbol{S}_{P_0}-\boldsymbol{F}_{P_0}(\boldsymbol{q}^{a}_{P_0}).
\end{equation}
Restrict $ \boldsymbol{q}^{a}_{P_0} $ from the first level to the second level as 
\begin{equation}
\boldsymbol{q}_{P_1}^{b} = \mathbb{P}_{P_1}^{P_0}\boldsymbol{q}_{P_0}^{a},
\end{equation}
where $ \boldsymbol{q}_{P_1}^{b}  $ is the initial solution at the second level, and $ \mathbb{P}_{P_1}^{P_0} $ indicates a projection from the first level (i.e., $P_0$) to the second level (i.e., $P_1$). 
Calculate the forcing term at this level as 
\begin{equation}
\boldsymbol{S}_{P_1} = \boldsymbol{F}_{P_1}(\boldsymbol{q}_{P_1}^{b}) + \mathbb{P}_{P_1}^{P_0} \boldsymbol{d}_{P_0}
\end{equation}
\item Smooth Eq.~\eqref{lvl_p_1} using the element Jacobi smoother Eq.~\eqref{smoothing_element_jacobi} for a few steps to obtain the smoothed solution $ \boldsymbol{q}_{P_1}^{a} $ at the second level. Update $ \boldsymbol{F} $ as $ \boldsymbol{F}_{P_1}(\boldsymbol{q}_{P_1}^{a}) $ afterwards.
\item The defect at the second level is 
\begin{equation}
\boldsymbol{d}_{P_1} = \boldsymbol{S}_{P_1}-\boldsymbol{F}_{P_1}(\boldsymbol{q}_{P_1}^a).
\end{equation}
Restrict the solution $ \boldsymbol{q}^{a}_{P_1}$ from the second level to the third level  as 
\begin{equation}
\boldsymbol{q}_{P_2}^{b} = \mathbb{P}_{P_2}^{P_1}\boldsymbol{q}_{P_1}^{a}.
\end{equation}
$ \boldsymbol{q}_{P_2}^{b}  $ is the initial solution at the third level. Calculate the forcing term at the third level as 
\begin{equation}
\boldsymbol{S}_{P_2} = \boldsymbol{F}_{P_2}(\boldsymbol{q}_{P_2}^{b}) +\mathbb{P}_{P_2}^{P_1} \boldsymbol{d}_{P_1} 
\end{equation}
\item Smooth Eq.~\eqref{lvl_p_2} using the element Jacobi smoother Eq.~\eqref{smoothing_element_jacobi} for a few steps to obtain the smoothed solution $ \boldsymbol{q}_{P_2}^{a} $ at the third level.
\item Correct the solution at the intermediate (second) level with a prolongation procedure as 
\begin{equation}
\boldsymbol{q}_{P_1}^{c,b}= \boldsymbol{q}_{P_1}^{a}+\mathbb{I}_{P_1}^{P_2}\boldsymbol{C}_{P_2},
\end{equation}
where $ \boldsymbol{C}_{P_2} = \boldsymbol{q}_{P_2}^{a}-\boldsymbol{q}_{P_2}^{b} $, and $ \mathbb{I}_{P_1}^{P_2}$ is an interpolation from the third level (i.e., $P_2$) to the second level (i.,e, $P_1$).
\item Post-smooth Eq.~\eqref{lvl_p_1} using the element Jacobi smoother Eq.~\eqref{smoothing_element_jacobi} for a few steps with starting value $ \boldsymbol{q}_{P_1}^{c,b} $  to obtain the smoothed solution $ \boldsymbol{q}_{P_1}^{c,a} $ at the second level.
\item Correct the solution at the finest (first) level as  
\begin{equation}
\boldsymbol{q}_{P_0}^{c,b}= \boldsymbol{q}_{P_0}^{a}+\mathbb{I}_{P_0}^{P_1}\boldsymbol{C}_{P_1}
\end{equation}
where $ \boldsymbol{C}_{P_1} = \boldsymbol{q}_{P_1}^{c,a}-\boldsymbol{q}_{P_1}^{b} $.
\item Post-smooth Eq.~\eqref{lvl_p_0} using the element Jacobi smoother Eq.~\eqref{smoothing_element_jacobi} for a few steps with starting value $ \boldsymbol{q}_{P_0}^{c,b} $ to obtain the smoothed solution $ \boldsymbol{q}_{P_0}^{c,a} $ at the finest level. $ \boldsymbol{q}_{P_0}^{c,a} $ is the final solution of $ \boldsymbol{q}_{P_0}$ after one V-cycle.
\end{itemize}
To simplify the notation, we use $ P\{P_0-P_1-P_2\} $ to denote a three-level V-cycle $ P $-multigrid solver, in which the hierarchy of the polynomial degrees is $ \{P_0-P_1-P_2-P_1-P_0\} $. The number of iterations of pre-smoothing and post-smoothing at the same level are identical. Therefore, we employ $ I\{n_{0}-n_{1}-n_{2}\} $ to denote the numbers of iterations in the smoothing procedure at different levels.
\section{Numerical results}\label{Results}

\subsection{Validation of spatiotemporal order of accuracy}
In this section, we employ the isentropic vortex propagation problem at low free stream Mach numbers to validate the order of accuracy for both spatial discretizations and time integrations. 
The free stream has the following flow conditions: $ (\rho, u, v, \text{Ma}_\infty) = (1.0, 1.0, 1.0, 0.005) $, and the fluctuation is defined as~\cite{bassi2015linearly}
\begin{equation}\label{VP_Solution}
\begin{cases}
\delta u=-\frac{\alpha}{2\pi}(y-y_0)e^{\phi(1-r^2)},\\
\delta v=\frac{\alpha}{2\pi}(x-x_0)e^{\phi(1-r^2)},\\
\delta T=-\frac{\alpha^2(\gamma-1)}{16\phi\gamma\pi^2}e^{2\phi(1-r^2)},\\
dS=0,\\
\end{cases}	
\end{equation}
where $ \phi = \frac{1}{2}$ and $ \alpha = 5 $ are parameters that define the vortex strength. $ r=(x-x_0)^{2}+(y-y_0)^{2} $ is the distance from any point $ (x,y) $ to the center of the vortex $ (x_0,y_0) = (0,0) $ at $ t=0 $. The periodic domain is defined in $ [-10,10]^2 $. 
At $ \text{Ma}_\infty = 0.005 $, the variation of the temperature $ T $ is trivial as well as the density $ \rho $ and the pressure $ p $. Therefore, this is a very good case to test the low dissipation and high resolution natures of high-order methods. 

Since there are no wall boundaries in this problem, the global cut-off is turned off. This indicates that $ \epsilon $ is set as the local Mach number instead of that in Eq.~\eqref{eps_nonmove}.
For the time refinement study, 
the $ P^5 $ (i.e., $6^{th}$ order) FR scheme is employed to solve the problem on a $ 50\times50  $ mesh to validate the order of accuracy for ESDIRK methods. We simulate this problem for one period, i.e., $ t_{end} = \widetilde{t} = 20 $.  A three-level V-cycle $ P $-multigrid solver of $ P\{5-3-1\} $ serves as the nonlinear solver. $ I\{20-20-40\} $ is adopted as the number of iterations for smoothing at different levels. We set $ \Delta \tau_{init} = 0.01 $ and $ \Delta \tau_{max} = 10 $ directly instead of providing the initial and maximum value of $ CFL $ for the pseudo transient continuation. The element Jacobi smoother is updated every 10 pseudo iterations. The convergence tolerance of the pseudo transient continuation is $ tol_{pseudo} = 10^{-4} $.  Numerical results of the time refinement study are presented in Table~\ref{vp_esdirk_convergence}. It is observed that all ESDIRK methods converge to the nominal order of accuracy except that order of reduction is observed for ESDIRK4 when the errors of $ \rho $ or $ p $ are considered.

The grid refinement study is conducted on a $ 12\times 12 $, $ 24\times24 $, and $ 36\times36$  mesh set. ESDIRK4 is used for the time integration and $ \Delta t = \widetilde{t}/200 $. We only simulate this problem for $ t_{end} = 2 $. The hierarchies of the polynomial degrees of the $ P $-multigrid solvers are $ P\{3-2-1\} $ and $ P\{4-2-1\} $  for $ P^3 $ and $ P^4 $ FR schemes, respectively. As shown in Table~\ref{vp_fr_convergence}, FR methods can preserve the nominal order of accuracy for velocity. However, there are order reductions for both $ \rho $ and $ p $ due to that the errors quickly drop to the accuracy limit of the solver as we refine the grids.
\begin{table}
\small
\centering
\caption{Time refinement study of the ESDIRK methods on solving the isentropic vortex propagation at $ \text{Ma}_\infty=0.005 $.} 
\label{vp_esdirk_convergence}
\begin{tabular}{rrrrrrrr}
\hline
\hline
&$ \Delta t $  & $ \rho $ & order & $ u  $& order & $ p $ & order\\
\hline
\multirow{3}{1.5cm}{ESDIRK2}
&$ \widetilde{t}/100 $ &3.6483e-07&     & 9.8315e-12&     & 4.6706e-08 & \\
&$ \widetilde{t}/200$ &9.3321e-08&1.97 & 2.4997e-12& 1.98& 1.1868e-08 & 1.98\\
&$ \widetilde{t}/400 $ &2.3385e-08&2.00 & 6.2593e-13& 2.00& 2.9651e-09 & 2.00\\	
\hline  
\hline	
\multirow{3}{1.5cm}{ESDIRK3}
&$ \widetilde{t}/100 $ &1.8968e-07&     & 3.0862e-12&     & 4.1661e-08 &\\
&$ \widetilde{t}/200$ &2.8674e-08&2.73 & 4.6512e-13& 2.73& 7.3317e-09 & 2.51\\
&$ \widetilde{t}/400$ &3.7789e-09&2.92 & 6.0674e-14& 2.94& 1.0117e-09 & 2.86\\
\hline
\hline
\multirow{3}{1.5cm}{ESDIRK4}
&$ \widetilde{t}/40 $ &9.3385e-08&     & 1.8414e-12&     & 1.3238e-08 &\\
&$ \widetilde{t}/80$ &6.6311e-09&3.82 & 1.2589e-13& 3.87& 5.6713e-09 & 3.73\\
&$ \widetilde{t}/120$ &4.8141e-09&0.79 & 2.5137e-14& 4.00&  9.9590e-10 & 1.39\\
\hline
\hline
\end{tabular}	
\end{table}

\begin{table}
\small
\centering
\caption{Grid refinement study of the FR methods on solving the isentropic vortex propagation at $ \text{Ma}_\infty=0.005 $.  $ L $ is the length of the periodic domain in the $ x $ or $ y $ direction. } 
\label{vp_fr_convergence}
\begin{tabular}{rrrrrrrr}
\hline
\hline
&$ \Delta x $  & $ \rho $ & order & $ u  $& order & $ p $ & order\\
\hline
\multirow{3}{1.5cm}{$ P^3 FR $}
&$ L/12 $ &3.4712e-08&    & 1.0335e-12&     & 1.6424e-08 & \\
&$ L/24 $ &3.2835e-09&3.40& 5.0024e-14& 4.37& 7.8227e-10&4.39 \\
&$ L/36 $ &2.6967e-09&0.49& 1.0237e-14& 3.91& 4.2379e-10 & 1.51\\	
\hline  
\hline	
\multirow{3}{1.5cm}{$ P^4 FR $}
&$ L/12 $ &7.5699e-09&     & 1.4991e-13&     & 3.3548e-09 &\\
&$ L/24 $ &1.6704e-09&2.18 & 5.6887e-15& 4.72& 2.2024e-10 & 3.93\\
&$ L/36 $ &1.6553e-09&0.02 & 1.0253e-15& 4.23& 1.9209e-10 & 0.34\\
\hline
\hline		
\end{tabular}	
\end{table}

\subsection{The impact of the hierarchy of polynomial degrees}
As aforementioned, 
we intend to study the impact of the hierarchy of polynomial degrees on the convergence speed of the two-level and three-level $ P $-multigrid solvers.  The FR schemes with $ P^3 $, $ P^4 $ and $ P^5 $ solution construction are tested in this section. For all $ P $-multigrid solvers, the polynomial degree at the lowest level is no less than one. 
For the $ P^3 $ FR discretization, possible configurations of the hierarchy of polynomial degrees  are $ P\{3-2\} $, $P\{3-1\}  $ and $ P\{3-2-1\} $.
For the $ P^4 $ FR discretization, possible setups  are $ P\{4-3\} $, $P\{4-2\}  $, $ P\{4-1\} $, $ P\{4-3-2\} $, $P\{4-3-1\} $ and $ P\{4-2-1\} $. The following combinations for the $ P^5 $ FR discretization are studied,  i.e.,  $ P\{5-3\} $, $ P\{5-2\} $, $P\{5-1\}$, $ P\{5-4-1\} $, $ P\{5-3-1\} $ and $ P\{5-2-1\} $. For all the studies in this section, if not specifically mentioned, $ CFL_{init} = 10^2 $, $ CFL_{max}=10^5 $ and $ I\{5-10-20\} $ is employed as the numbers of iterations for the smoothing procedure at different levels. 

\subsubsection{Inviscid flow over a NACA0012 airfoil} \label{subsec:invNACA0012}
\begin{figure}		
	\centering
	\begin{tabular}{cc}
		\includegraphics[width=6cm]{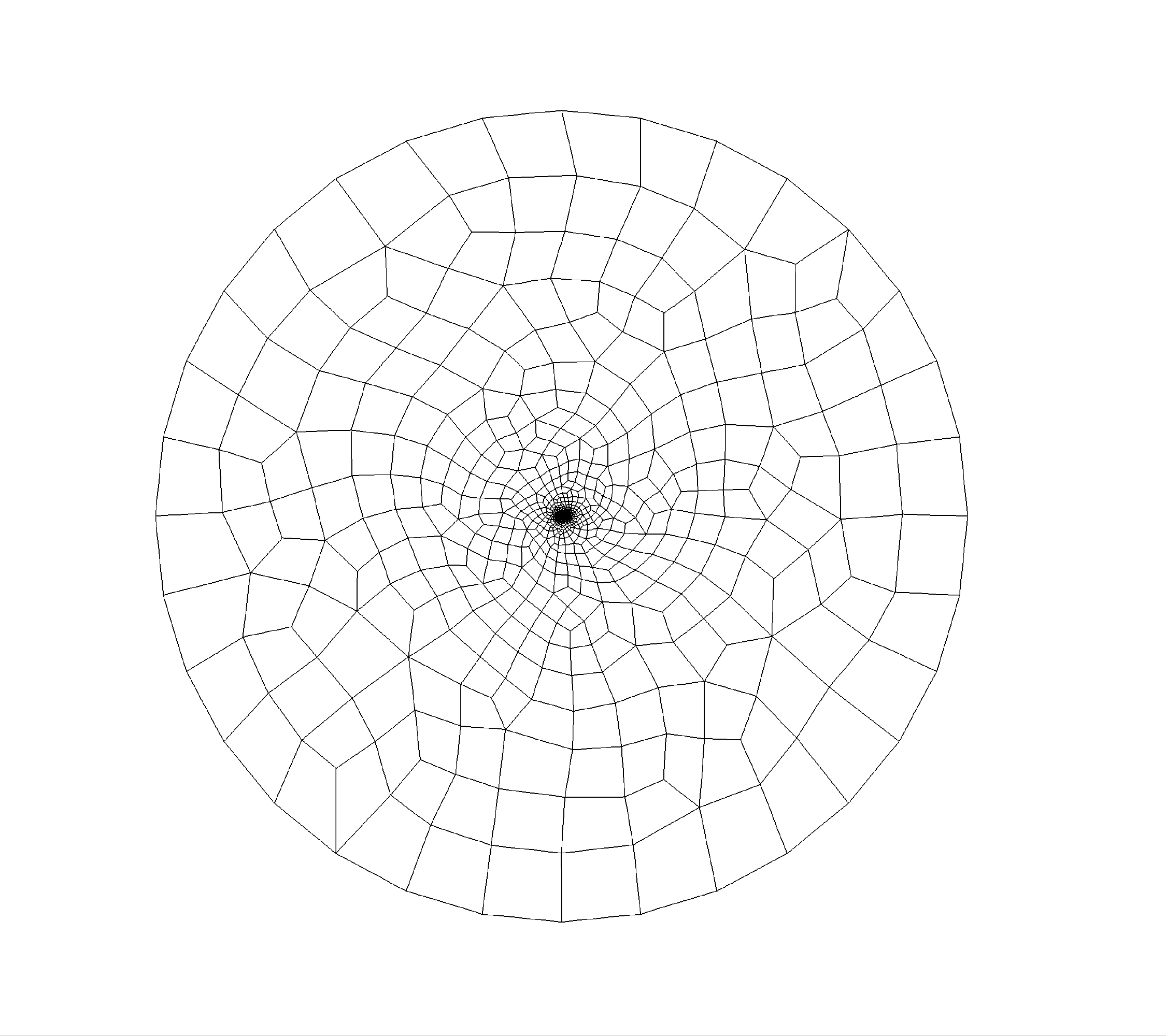} &
		\includegraphics[width=6cm]{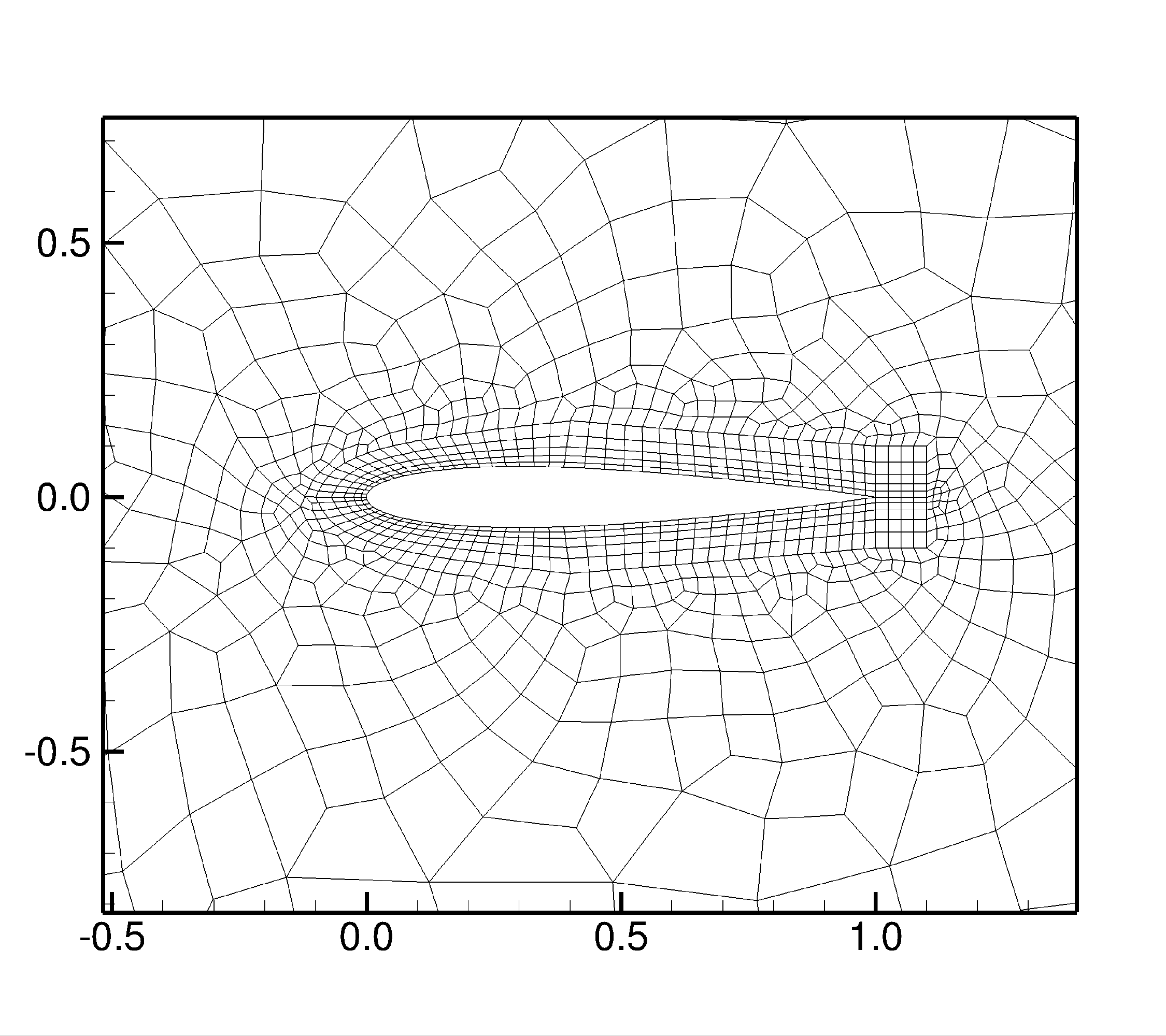}\\
		(a)& (b) \\	
	\end{tabular}
	\caption{Unstructured meshes around a NACA0012 airfoil. (a) A global view and (b) a close-up view near the airfoil.}	
	\label{naca_mesh}	
\end{figure}

We first simulate the inviscid flow with $ \text{Ma}=0.001 $ over a NACA0012 airfoil. The mesh that has 1560 quadrilateral elements is presented in Figure~\ref{naca_mesh}. The curved wall boundary is represented by $ P^4 $ elements. Note that the mesh is clustered near the wall to facilitate viscous simulation that will be presented in the next subsection.  We use $ I\{5-10-20\} $ for the smoothing procedure in the three-level $ P $-multigrid solver, $ I\{5-10\} $ for the two-level $ P $-multigrid solver and $ I\{10\} $ for the single-level solver.

\begin{figure}		
	\centering
	\begin{tabular}{cc}
		\includegraphics[width=6cm]{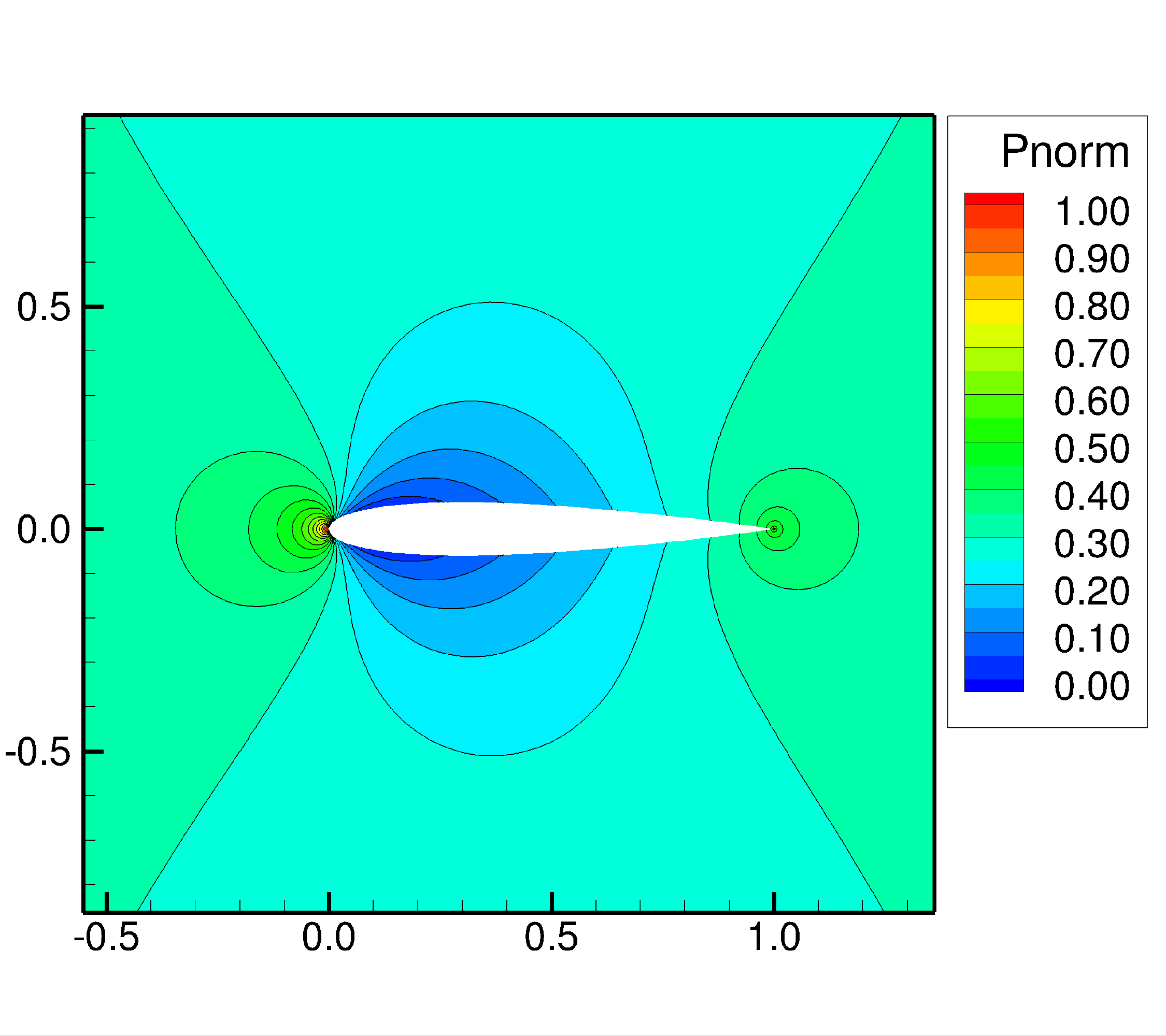} &
		\includegraphics[width=6cm]{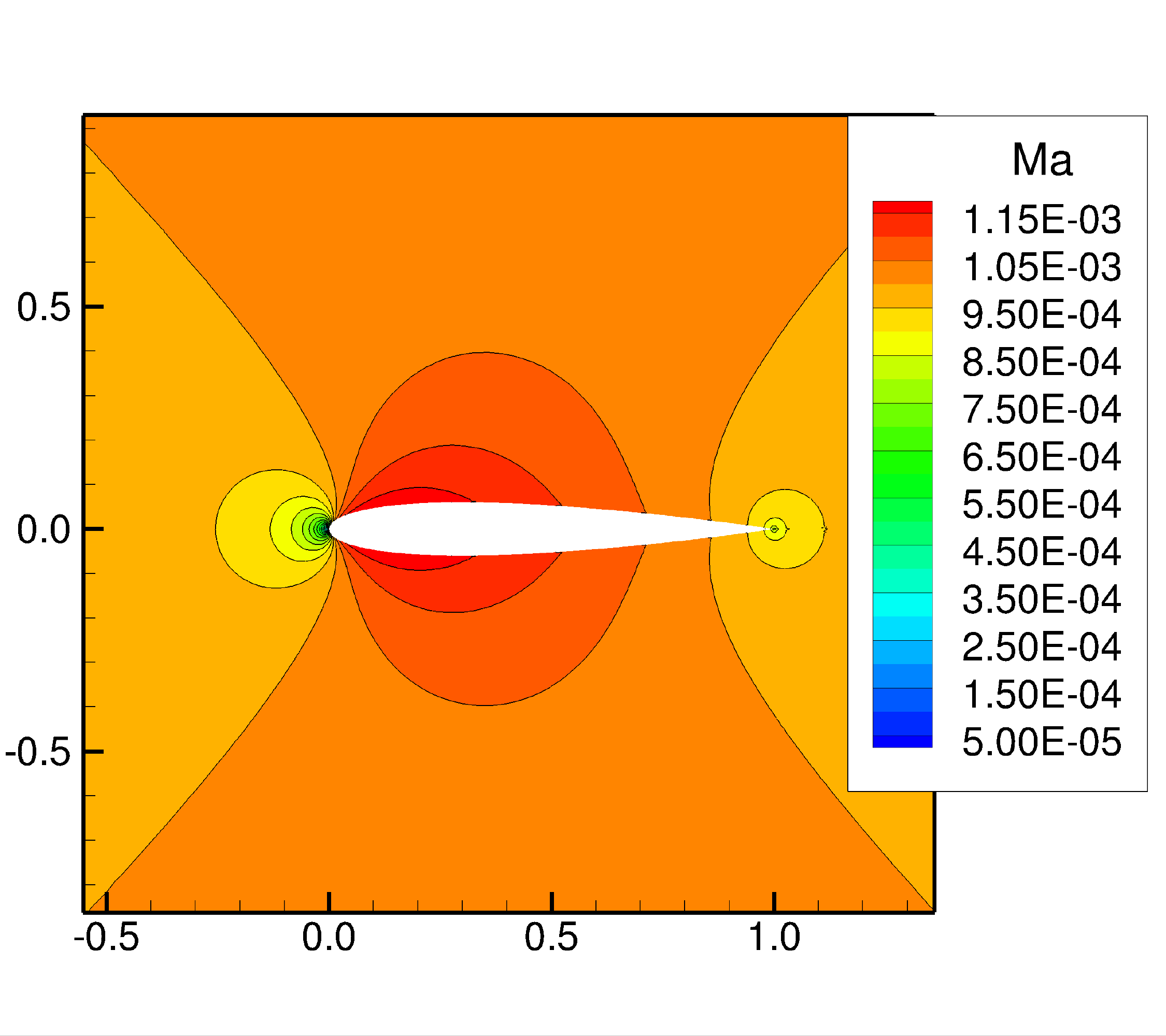}\\
		(a)& (b) \\	
	\end{tabular}
	\caption{(a)  Normalized pressure field and (b) Ma number field of the inviscid flow over a NACA0012 airfoil at $ \text{Ma}=0.001 $.}	
	\label{naca_contour}	
\end{figure}

\begin{figure}		
	\centering
	\begin{tabular}{cc}
		\includegraphics[width=6cm]{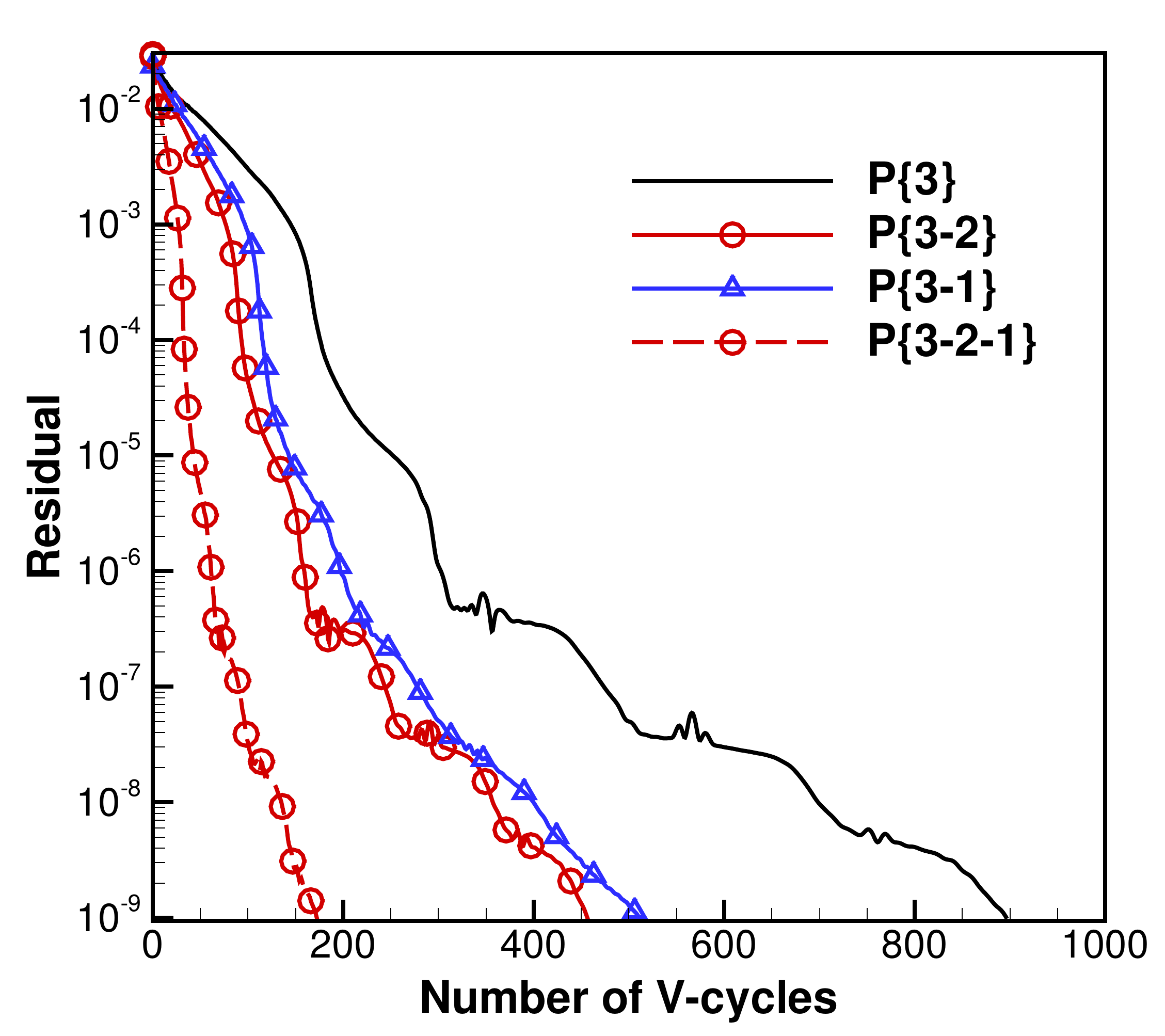} &
		\includegraphics[width=6cm]{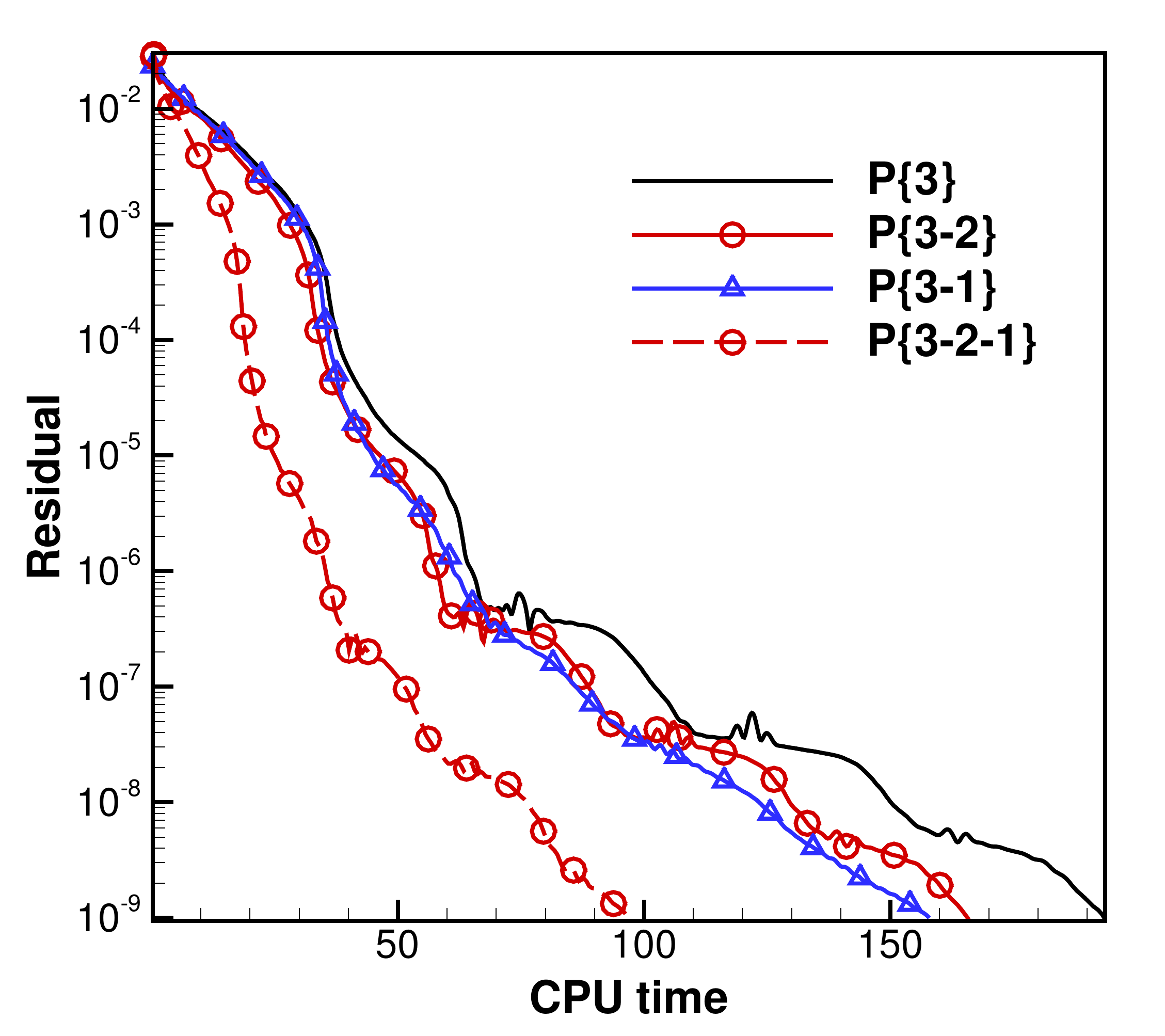}\\
		(a)& (b) \\	
	\end{tabular}
	\caption{Convergence histories of different $ P $-multigrid solvers for the $ P^3 $ FR discretization when solving the inviscid flow  over a NACA0012 airfoil  at $ \text{Ma}=0.001 $.}	
	\label{naca_inv_p3}	
\end{figure}

\begin{figure}		
	\centering
	\begin{tabular}{cc}
		\includegraphics[width=6cm]{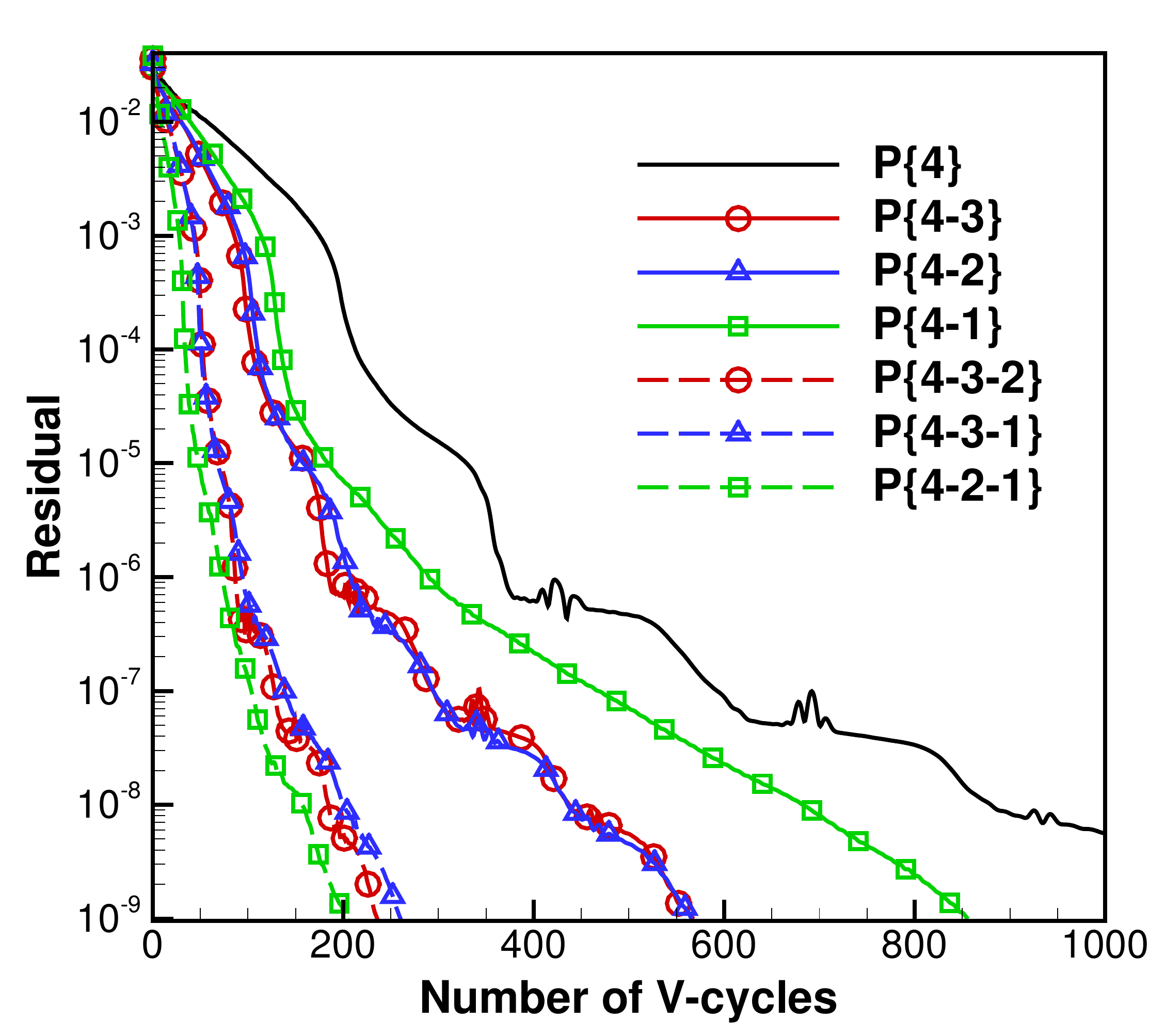} &
		\includegraphics[width=6cm]{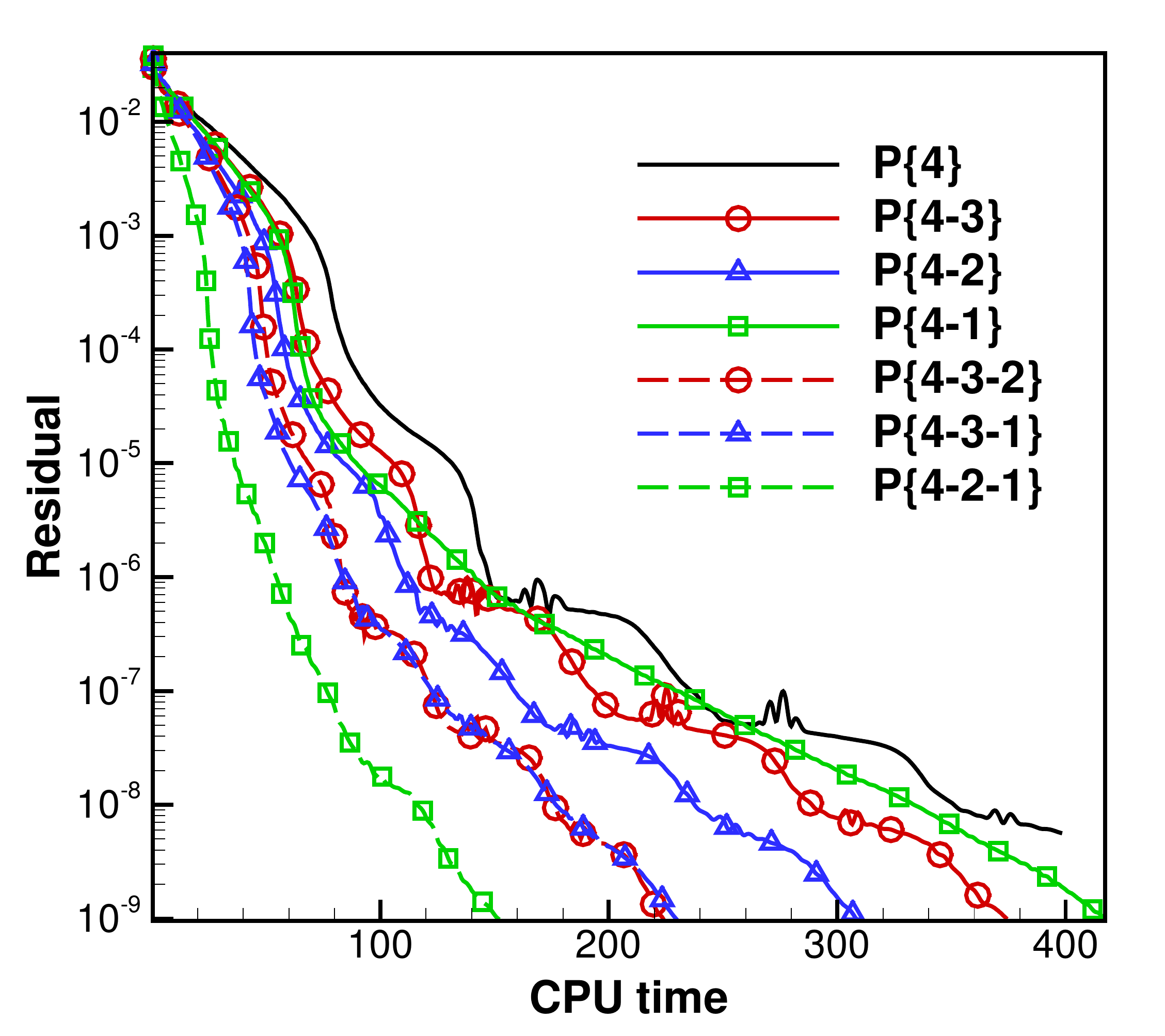}\\
		(a)& (b) \\	
	\end{tabular}
	\caption{Convergence histories  of different $ P $-multigrid solvers for the $ P^4 $ FR discretization when solving the inviscid flow  over a NACA0012 airfoil at $ \text{Ma}=0.001 $.}	
	\label{naca_inv_p4}	
\end{figure}

\begin{figure}		
	\centering
	\begin{tabular}{cc}
		\includegraphics[width=6cm]{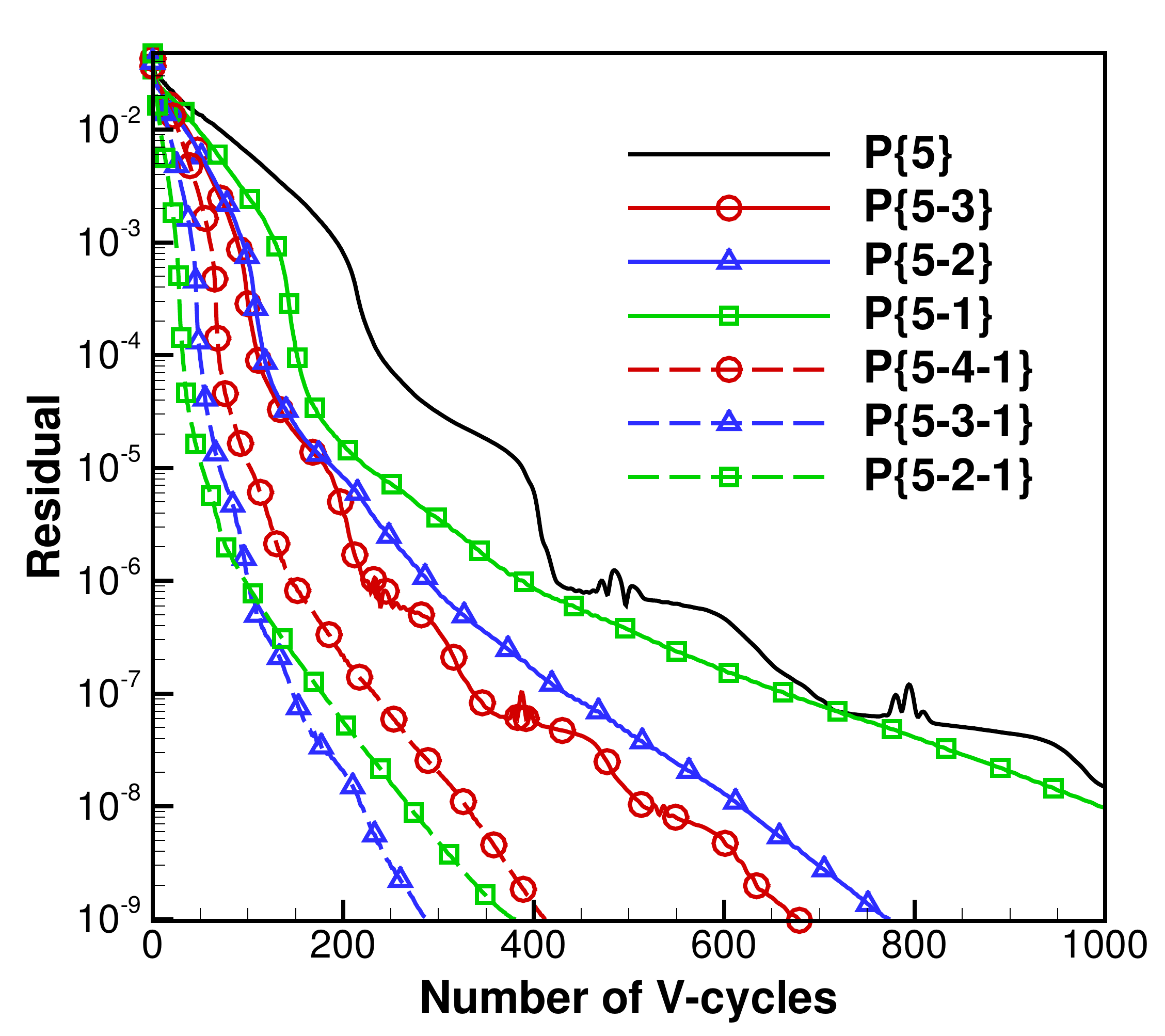} &
		\includegraphics[width=6cm]{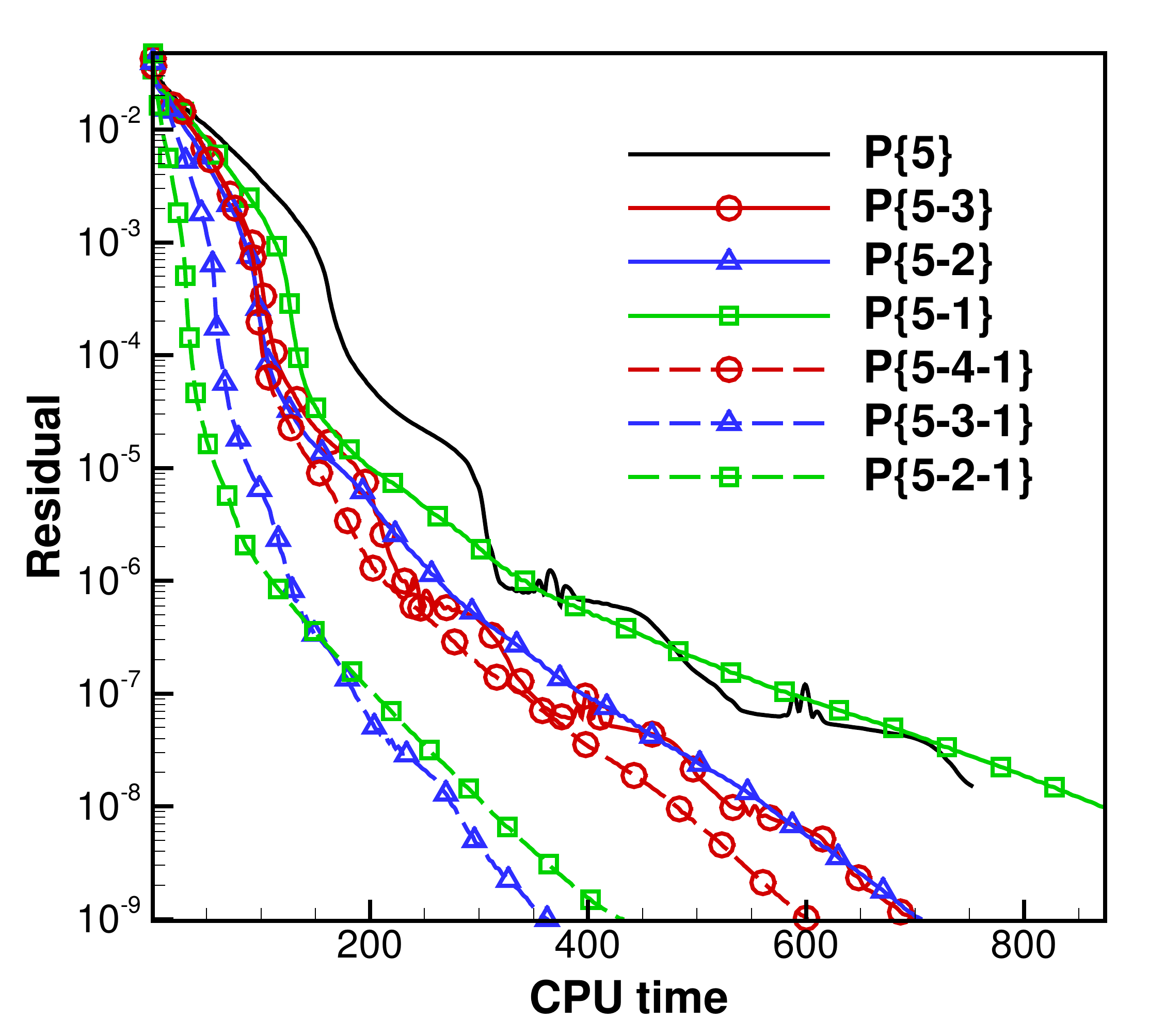}\\
		(a)& (b) \\	
	\end{tabular}
	\caption{Convergence histories of different $ P $-multigrid solvers for the $ P^5 $ FR discretization when solving the inviscid flow  over a NACA0012 airfoil at $ \text{Ma}=0.001 $.}	
	\label{naca_inv_p5}	
\end{figure}

We present the fields of the normalized pressure $ p_{norm} =\frac{p-p_{min}}{p_{max}-p_{min}}$ and Mach number in Figure~\ref{naca_contour}. We observe that no pressure oscillations occur near the stagnation point on the leading edge.
Convergence histories are present in Figure~\ref{naca_inv_p3}, Figure~\ref{naca_inv_p4} and Figure~\ref{naca_inv_p5}. The convergence performance with the $ P $-multigrid method has a significant improvement over that with a single level iterative method in terms of both CPU time and number of V-cycles, especially when three-level methods are employed. 
In general, a three-level V-cycle $ P $-multigrid method converges faster than a two-level V-cycle method. 

When examining the convergence histories of the two-level $ P $-multigrid methods carefully, we find that if the hierarchy of polynomial degrees deviate from $ \{P_0-P_0/2-P_0\} $, the performance of the $ P $-multigrid solver will get worse. For the $ P^3 $ FR method, $ P\{3-2\} $ and $ P\{3-1\} $ have almost the same computational cost. However, when the polynomial degree increases, $ P\{4-1\} $ and $ P\{5-1\} $ have the worst convergence speed compared to their counterparts. Additionally, $ P\{4-1\} $ and $ P\{5-1\} $ perform better than a single level method when the residual is above $ 10^{-6} $. However, when the residual further decreases, almost no acceleration can be gained from these two $ P $-multigrid solvers.   On the contrary, $ P\{4-2\} $ has the fastest convergence speed in all the two-level methods in Figure~\ref{naca_inv_p4}; for the $ P^5 $ FR method, $ P\{5-3\} $ and $ P\{5-2\} $ have almost the same convergence speed. The above observations suggest that  when the difference of the polynomial degrees between two adjacent levels is excessively large, the correction on the finer level from the coarser level becomes less effective when the residual becomes smaller, and the ineffective correction can even deteriorate the convergence rate. Similarly, for the three-level V-cycle $ P $-multigrid method, $ P\{4-2-1\} $ has the best performance in Figure~\ref{naca_inv_p4}; the convergence performance of $ P\{5-3-1\} $ and $ P\{5-2-1\} $ are close to each other and both better than that of $ P\{5-4-1\} $.  In summary, to achieve better convergence performance, the difference of the polynomial degrees between two adjacent levels should be close to half of the polynomial degree at the finer level.

\subsubsection{Viscous flow over a NACA0012 airfoil}
The viscous flow over a NACA0012 airfoil at $ \text{Ma}=0.001 $ and $ \text{Re}=5000 $ is studied in this subsection. We use the same mesh as that presented in Figure~\ref{naca_mesh}.

\begin{figure}		
	\centering
	\begin{tabular}{cc}
		\includegraphics[width=6cm]{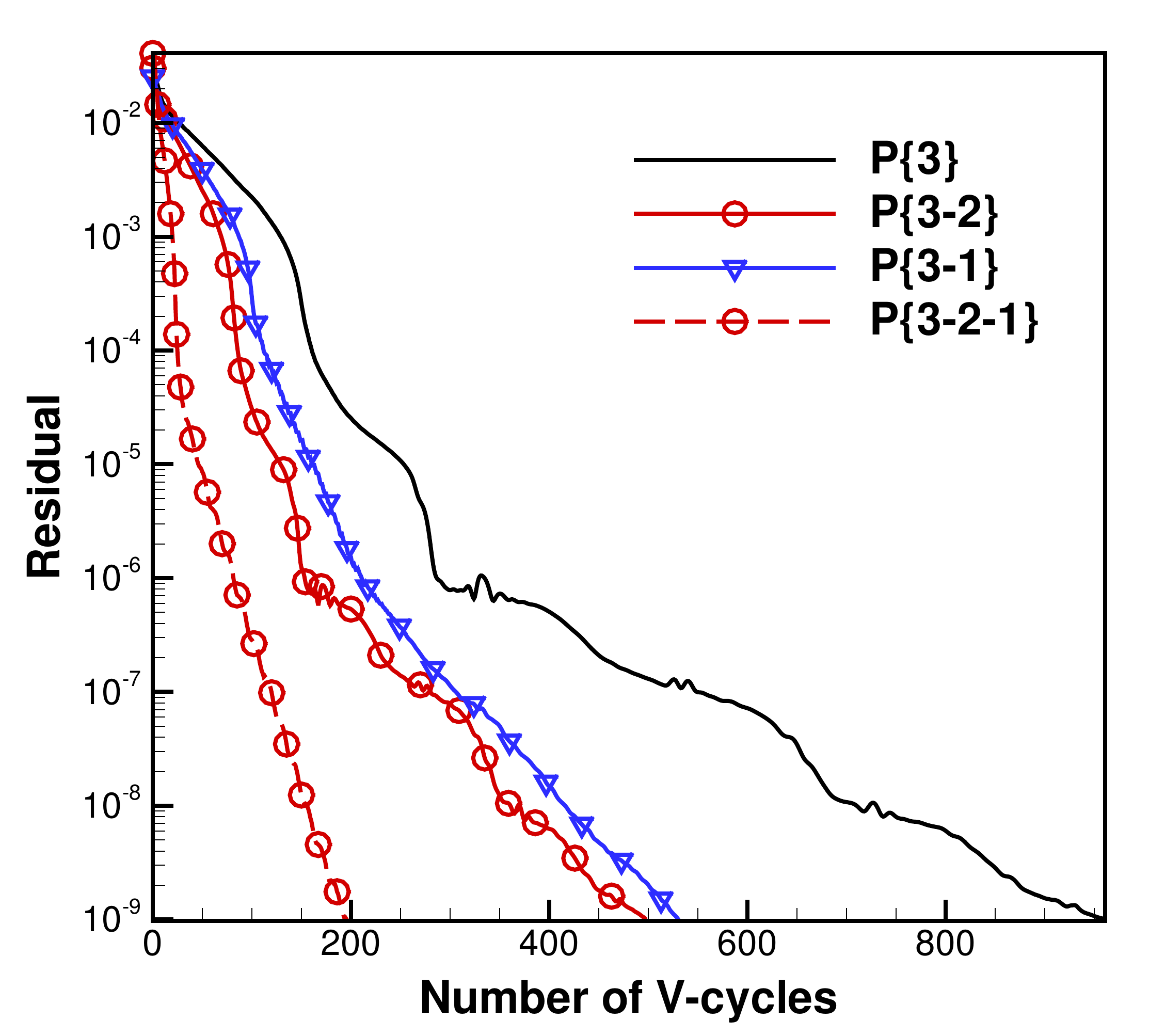} &
		\includegraphics[width=6cm]{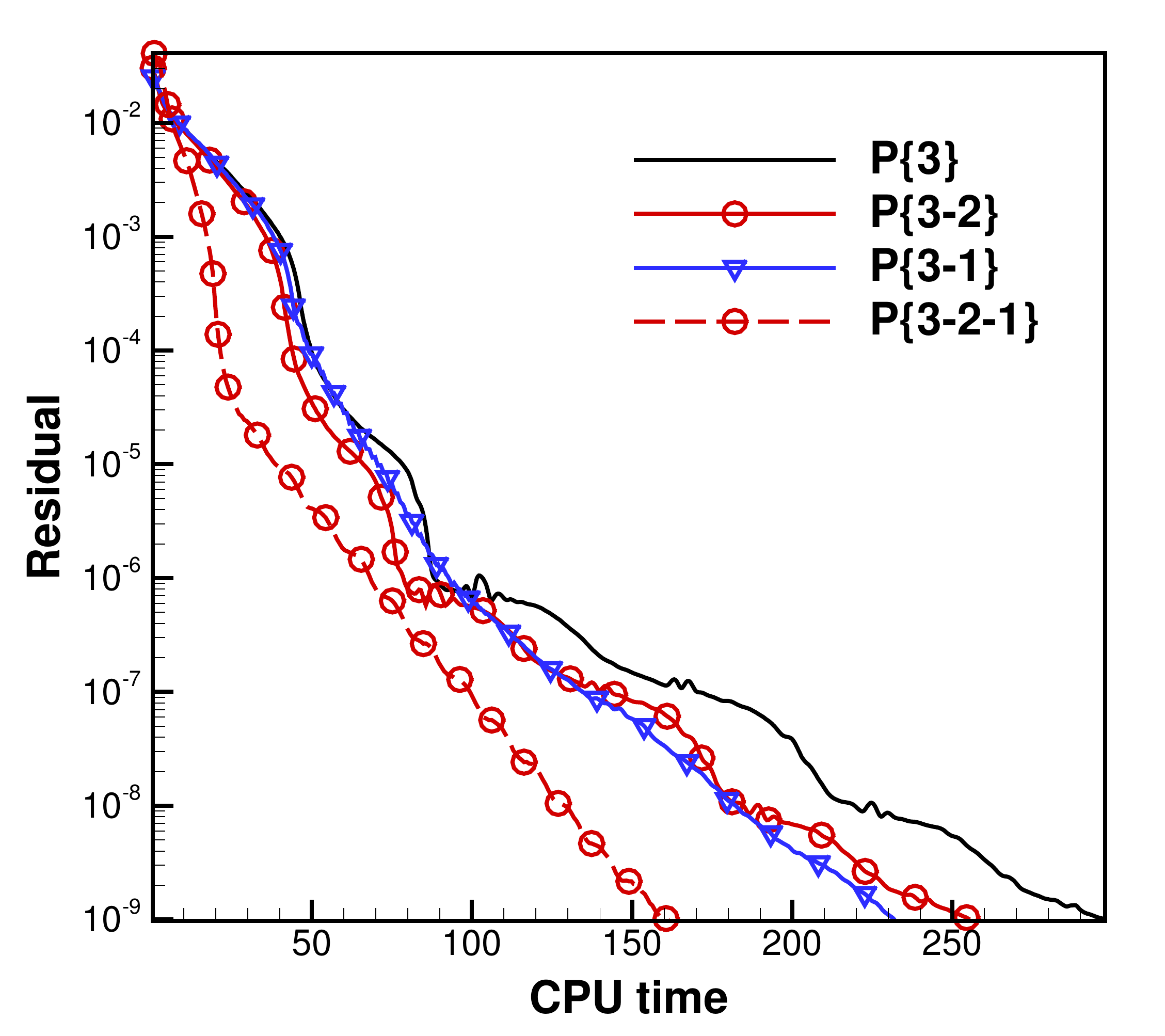}\\
		(a)& (b) \\	
	\end{tabular}
	\caption{Convergence histories of different $ P $-multigrid solvers for the $ P^3 $ FR discretization when solving viscous flow over a NACA0012 airfoil at $ \text{Ma}=0.001 $ and $ \text{Re}=5000 $.}	
	\label{naca_vis_p3}	
\end{figure}

\begin{figure}		
	\centering
	\begin{tabular}{cc}
		\includegraphics[width=6cm]{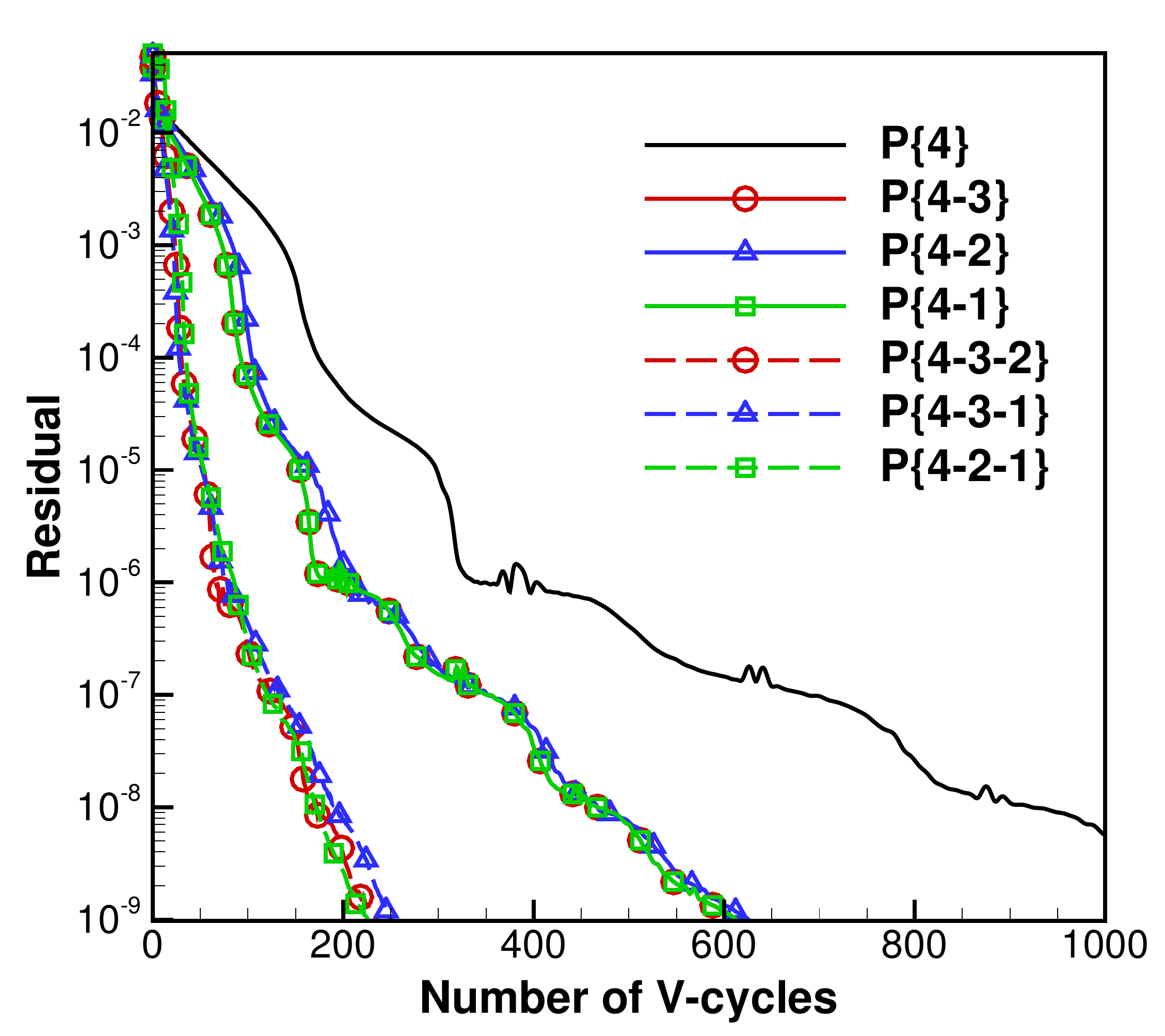} &
		\includegraphics[width=6cm]{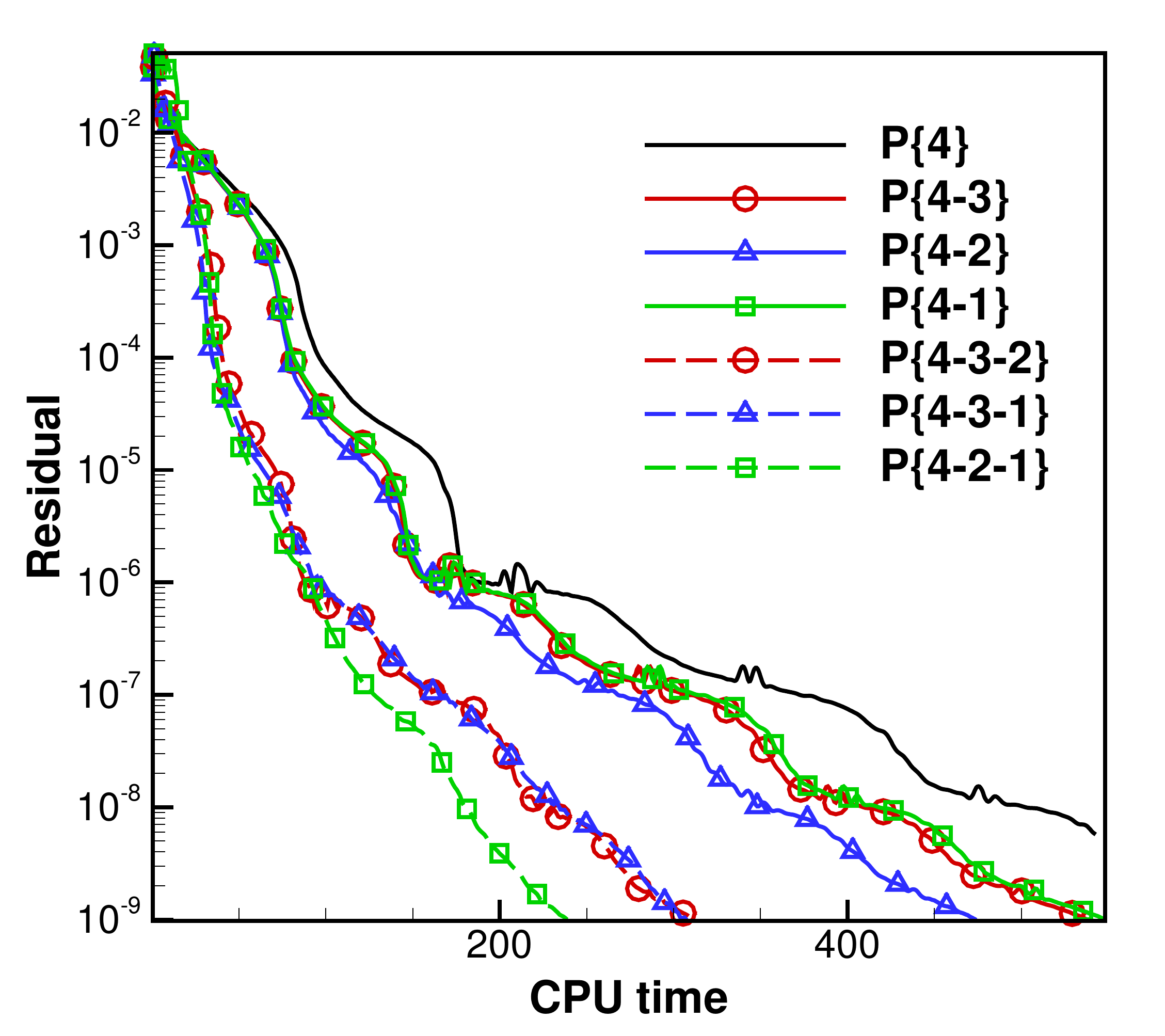}\\
		(a)& (b) \\	
	\end{tabular}
	\caption{Convergence histories  of different $ P $-multigrid solvers for the $ P^4 $ FR discretization when solving the viscous  flow over a NACA0012 airfoil at $ \text{Ma}=0.001 $ and $ \text{Re}=5000 $.}	
	\label{naca_vis_p4}	
\end{figure}

\begin{figure}		
	\centering
	\begin{tabular}{cc}
		\includegraphics[width=6cm]{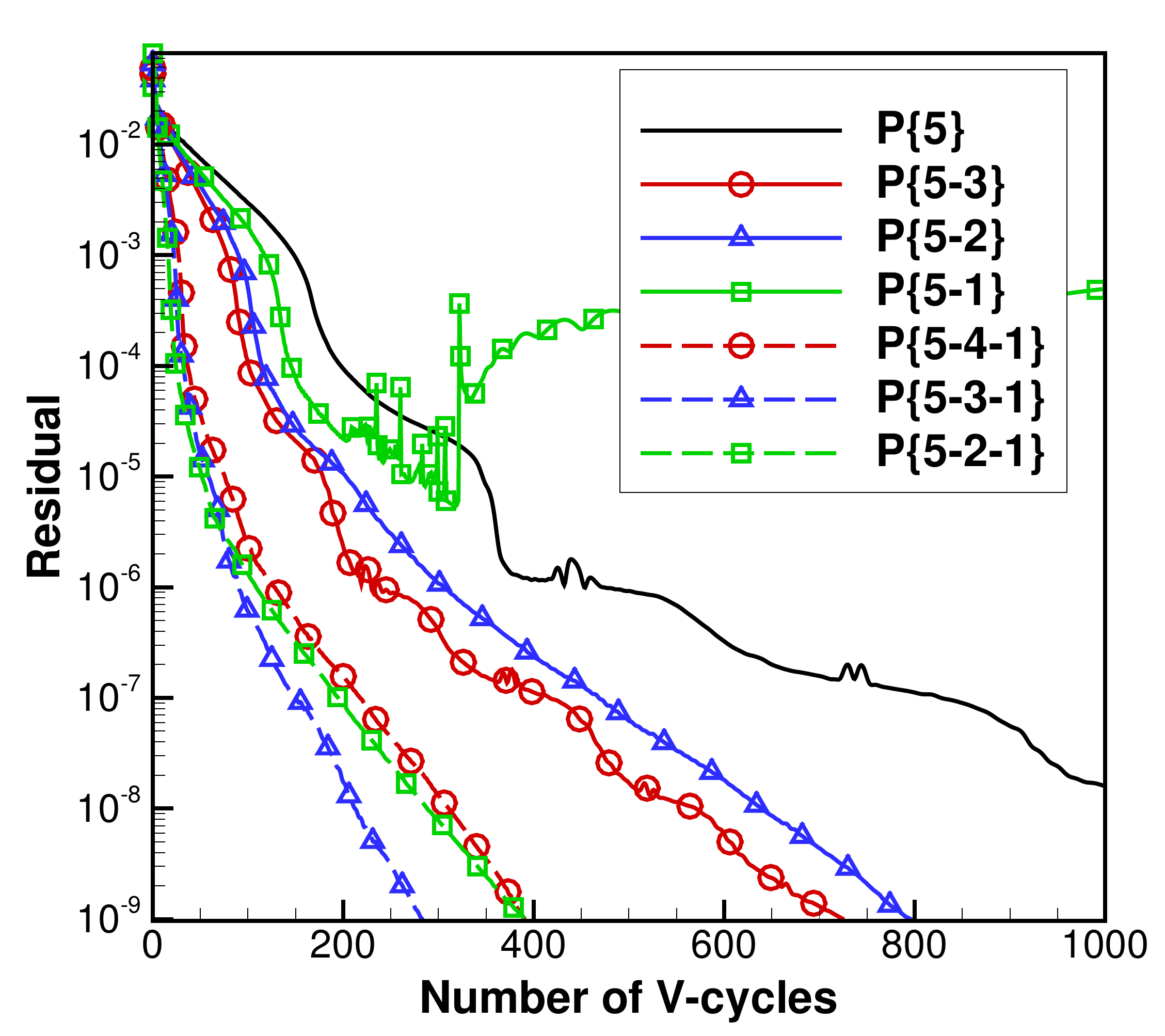} &
		\includegraphics[width=6cm]{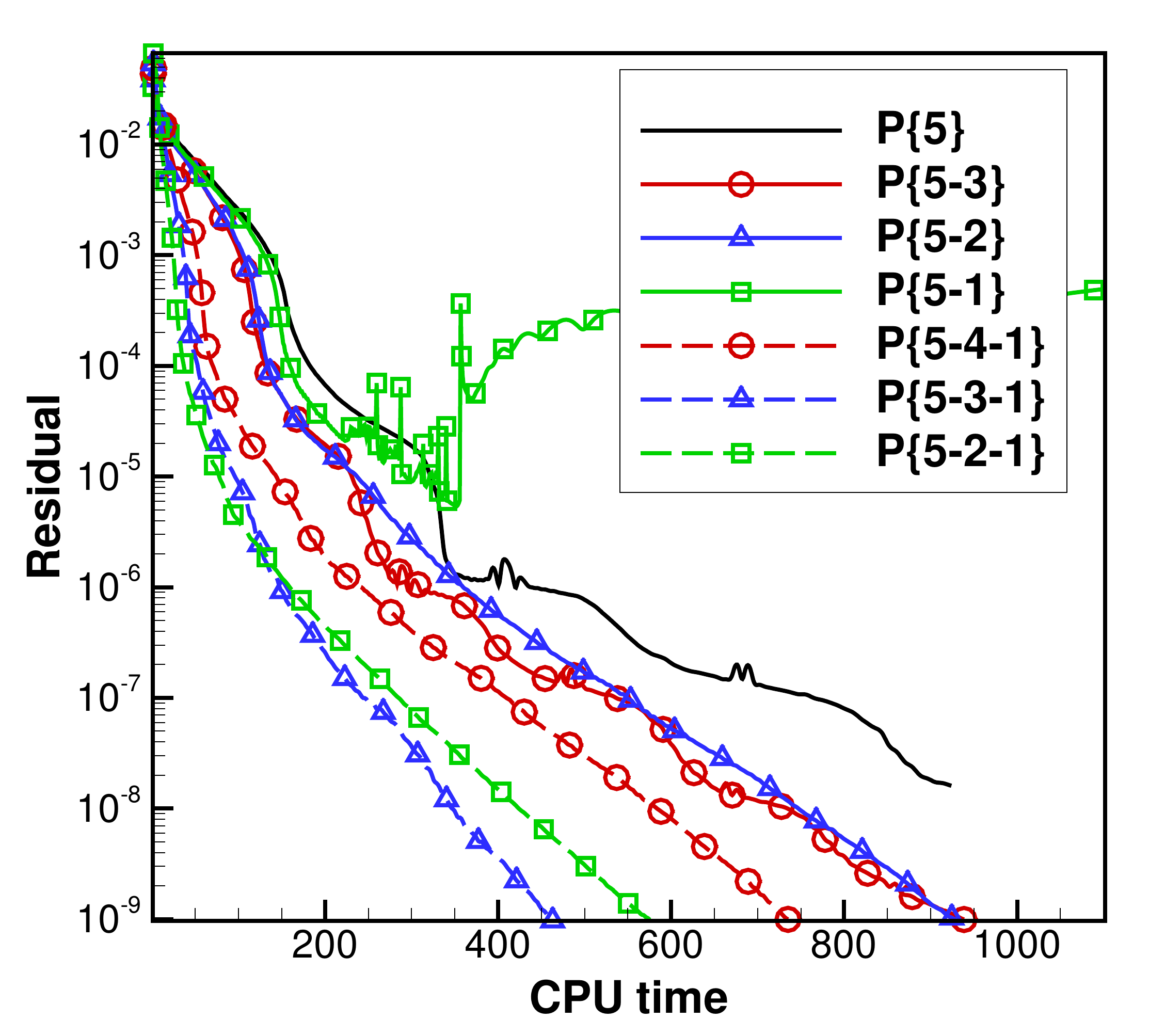}\\
		(a)& (b) \\	
	\end{tabular}
	\caption{Convergence histories of different $ P $-multigrid solvers for the $ P^5 $ FR discretization when solving the viscous flow over a NACA0012 airfoil at $ \text{Ma}=0.001 $ and $ \text{Re}=5000 $.}	
	\label{naca_vis_p5}	
\end{figure}

The convergence histories of different $ P $-multigrid solvers are presented in Figure~\ref{naca_vis_p3}, Figure~\ref{naca_vis_p4}  and Figure~\ref{naca_vis_p5}. Similar to the observations in Section~\ref{subsec:invNACA0012}, for two-level methods, $ P\{3-1\} $ and $P\{3-2\}$ have a similar performance in terms of both CPU time and number of V-cycles; $ P\{4-2\} $ is the best in all two-level methods for the $ P^4 $ FR discretization; the convergence performance of $ P\{5-3\} $ and $ P\{5-2\} $ is close to each other. For  $ P\{5-1\} $, the residual starts to oscillate after it drops below $ 10^{-4} $ and fails to converge. This indicates that the `correction' from the coarser level no longer favors the convergence at the finer level, and even worse, it introduces new errors that lead to failure of further convergence. For three-level methods, 
we observe that the $ P $-multigrid solvers which adopt a polynomial degree hierarchy close to $ \{P_0-P_0/2-P_0/4-P_0/2-P_0\} $ tend to have the best performance. This is consistent with the observation from numerical experiments for the inviscid flows.

\subsubsection{Inviscid flow over a sphere}
\begin{figure}		
\centering
\begin{tabular}{ccc}
\includegraphics[width=4cm]{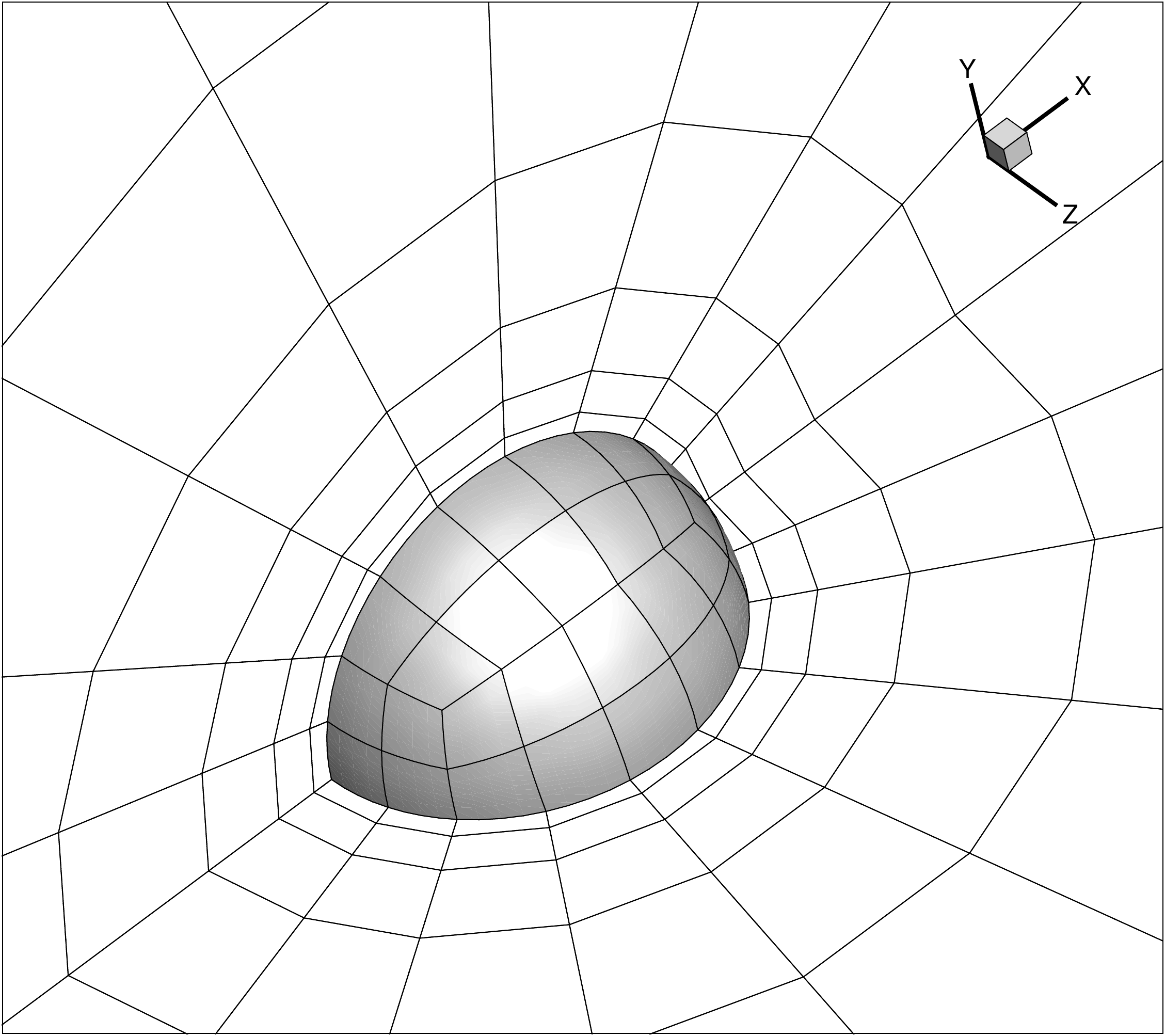} &
\includegraphics[width=4cm]{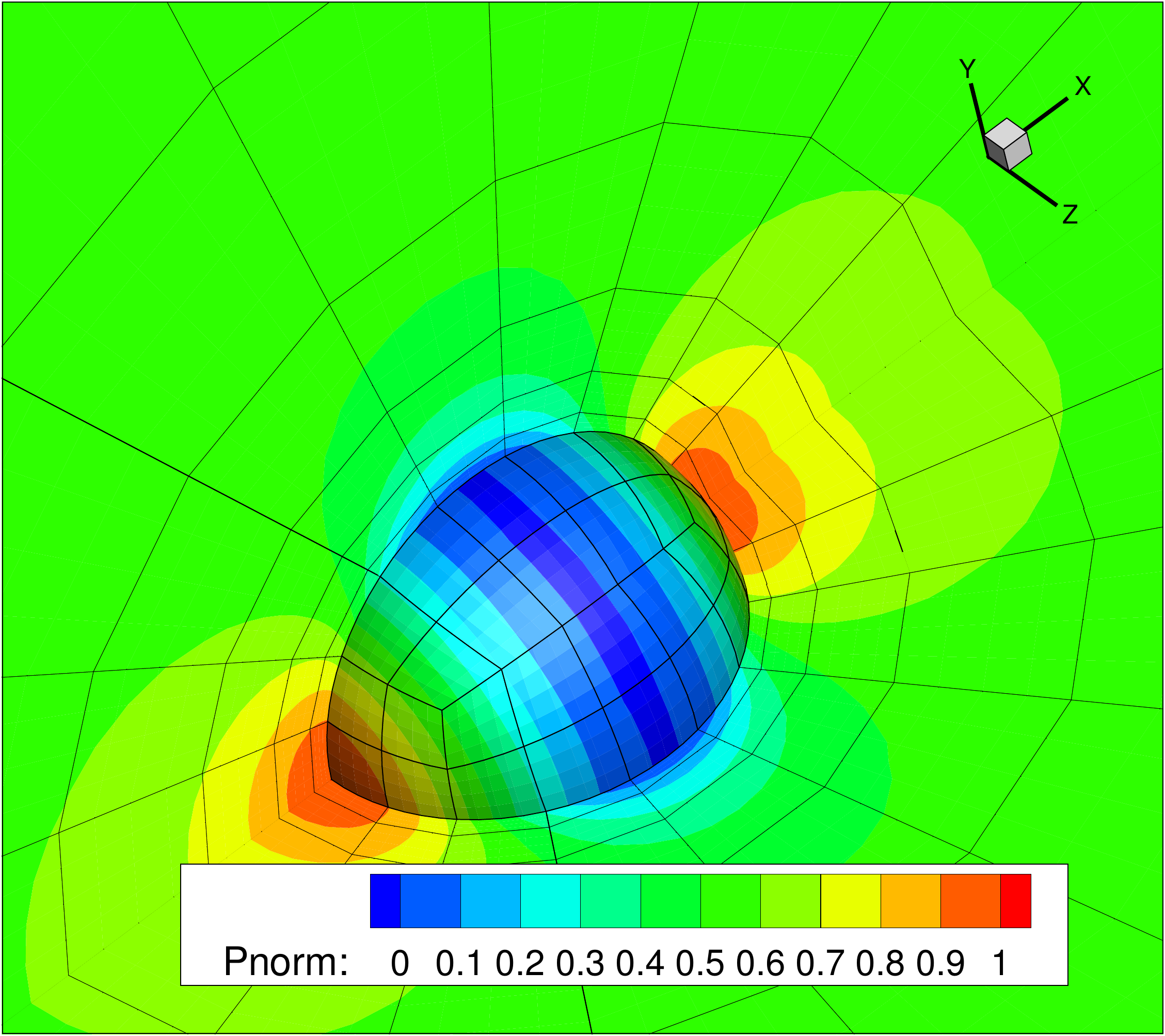}&
\includegraphics[width=4cm]{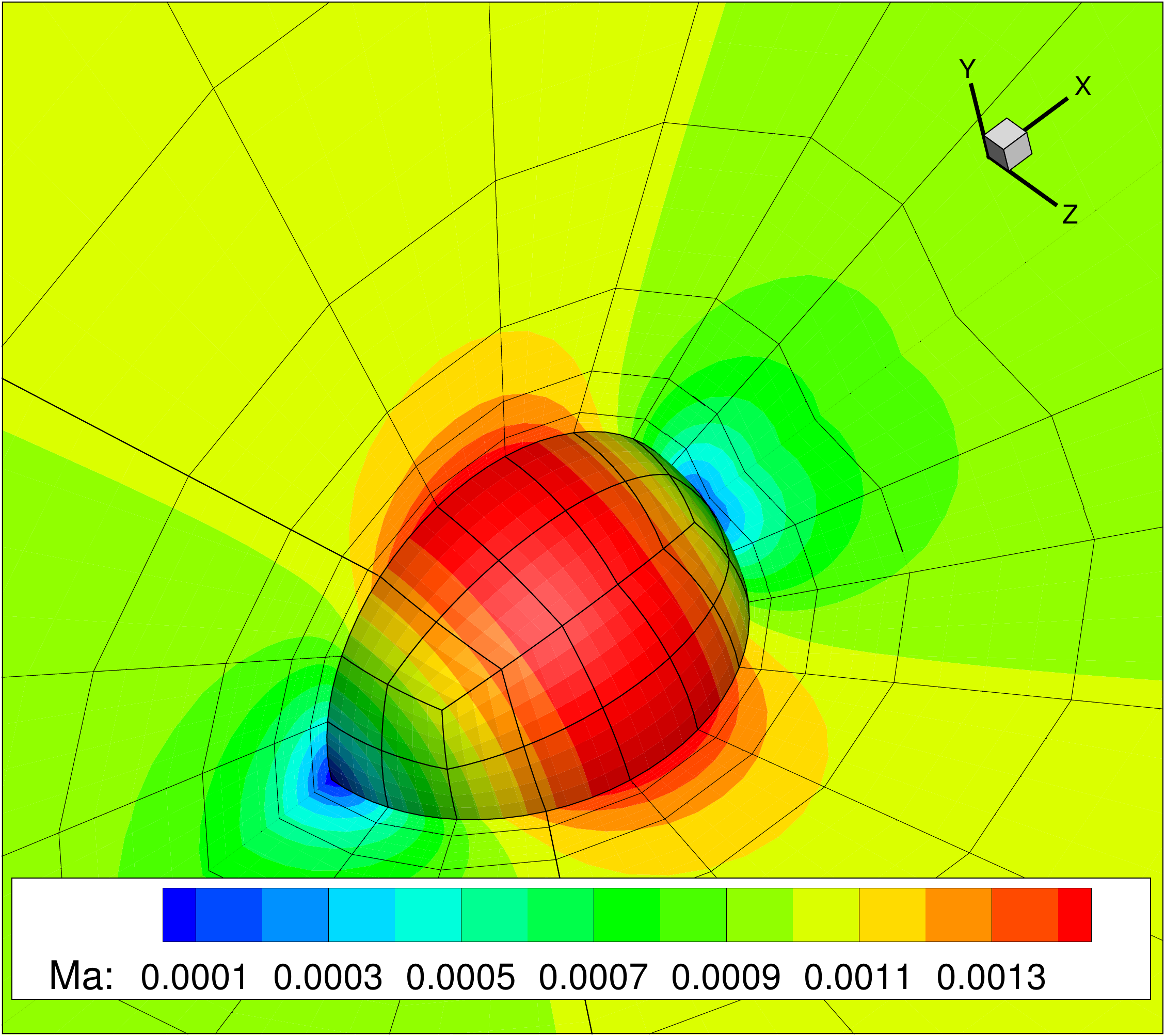}
\\
(a)& (b)&(c) \\	
\end{tabular}
\caption{Inviscid flow over a sphere at $ \text{Ma} =0.001 $. (a) Meshes in the near-wall region, (b) contour of the normalized pressure $ p_{norm} $ and (c)  contour of the Mach number.}
\label{sphere_inviscid}	
\end{figure}

\begin{figure}		
	\centering
	\begin{tabular}{cc}
		\includegraphics[width=6cm]{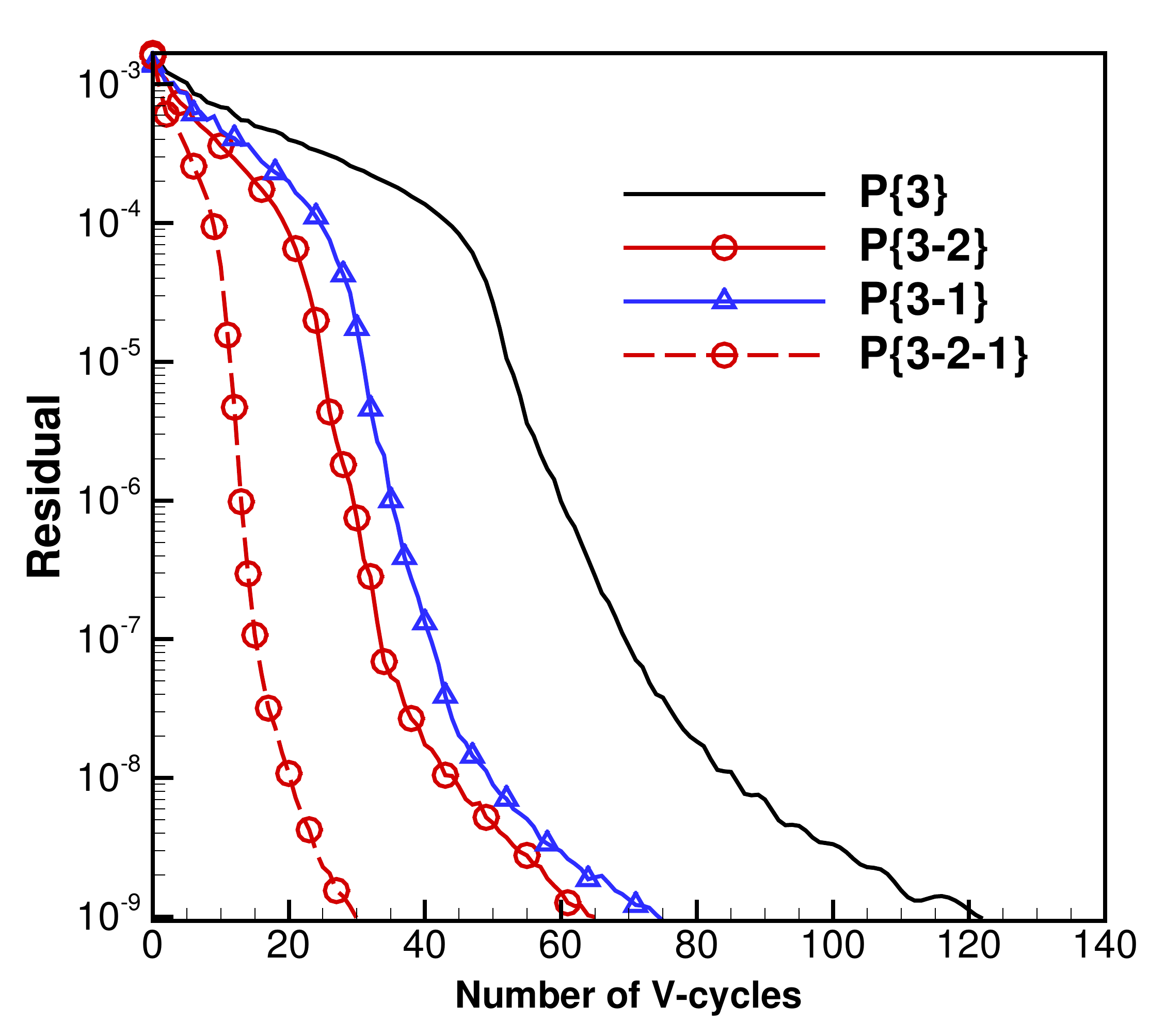} &
		\includegraphics[width=6cm]{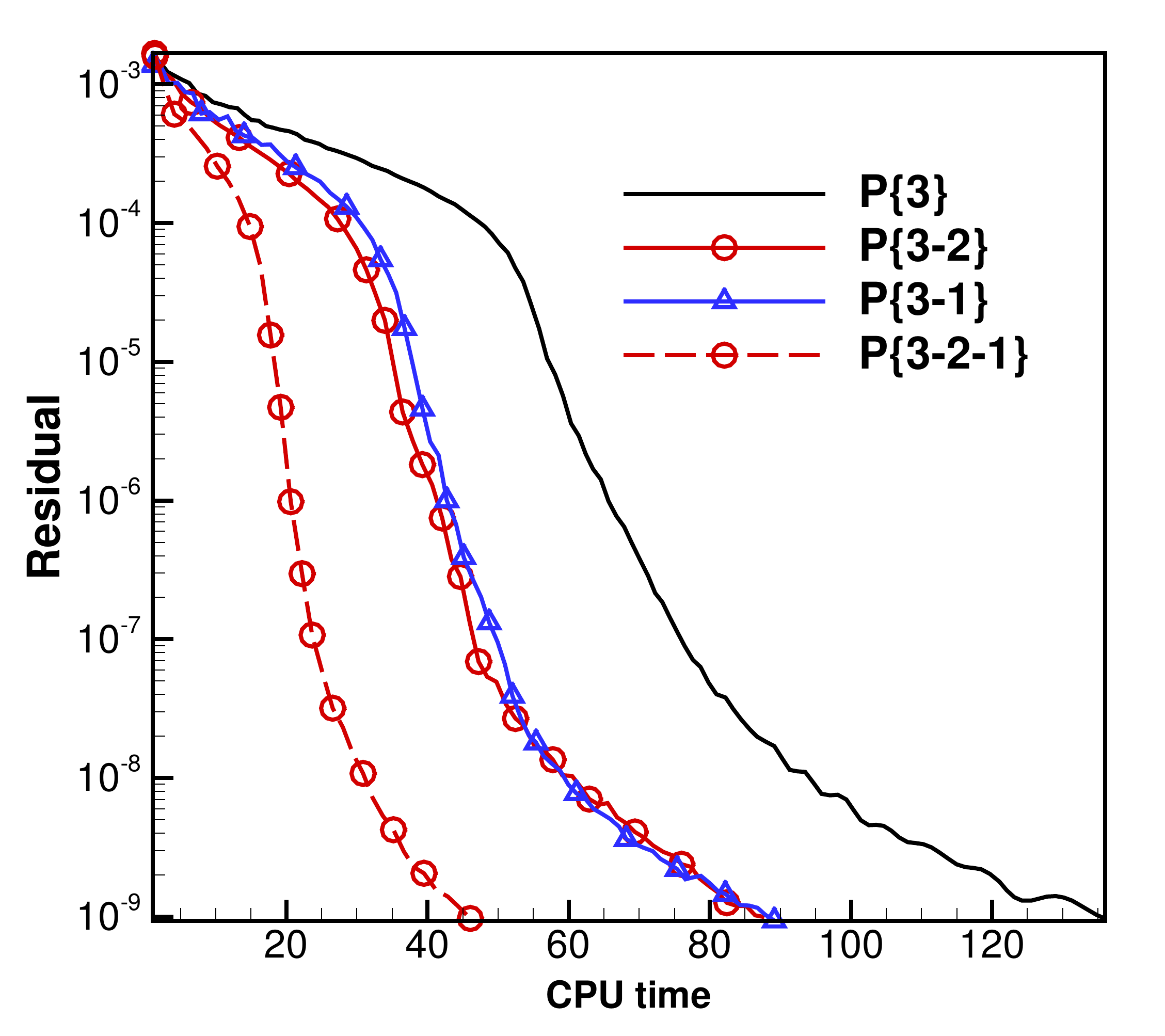}\\
		(a)& (b) \\	
	\end{tabular}
	\caption{Convergence histories of different $ P $-multigrid solvers for the $ P^3 $ FR discretization when solving the inviscid flow over a sphere at $ \text{Ma}=0.001 $.}	
	\label{sphere_inv_p3}	
\end{figure}

\begin{figure}		
	\centering
	\begin{tabular}{cc}
		\includegraphics[width=6cm]{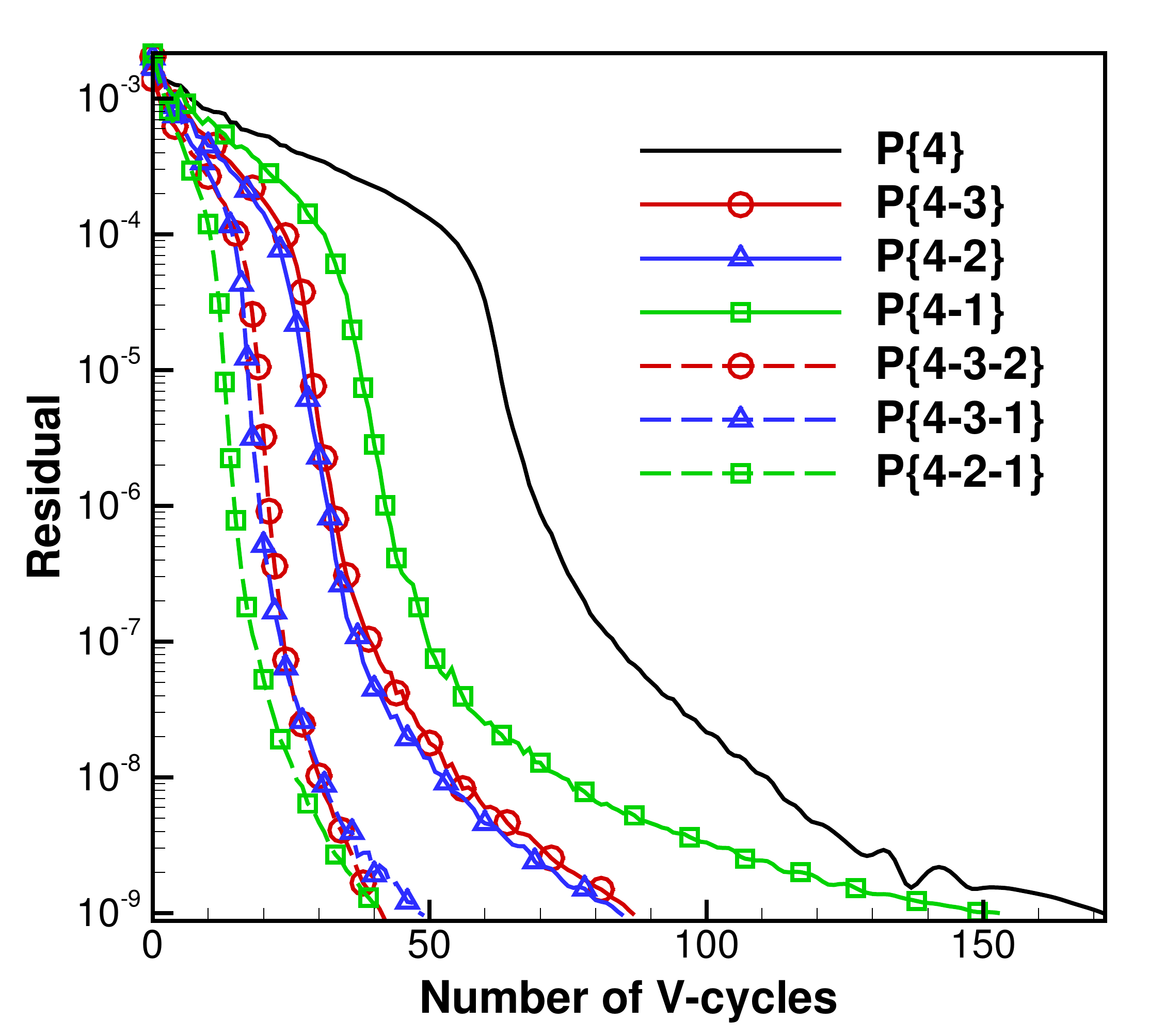} &
		\includegraphics[width=6cm]{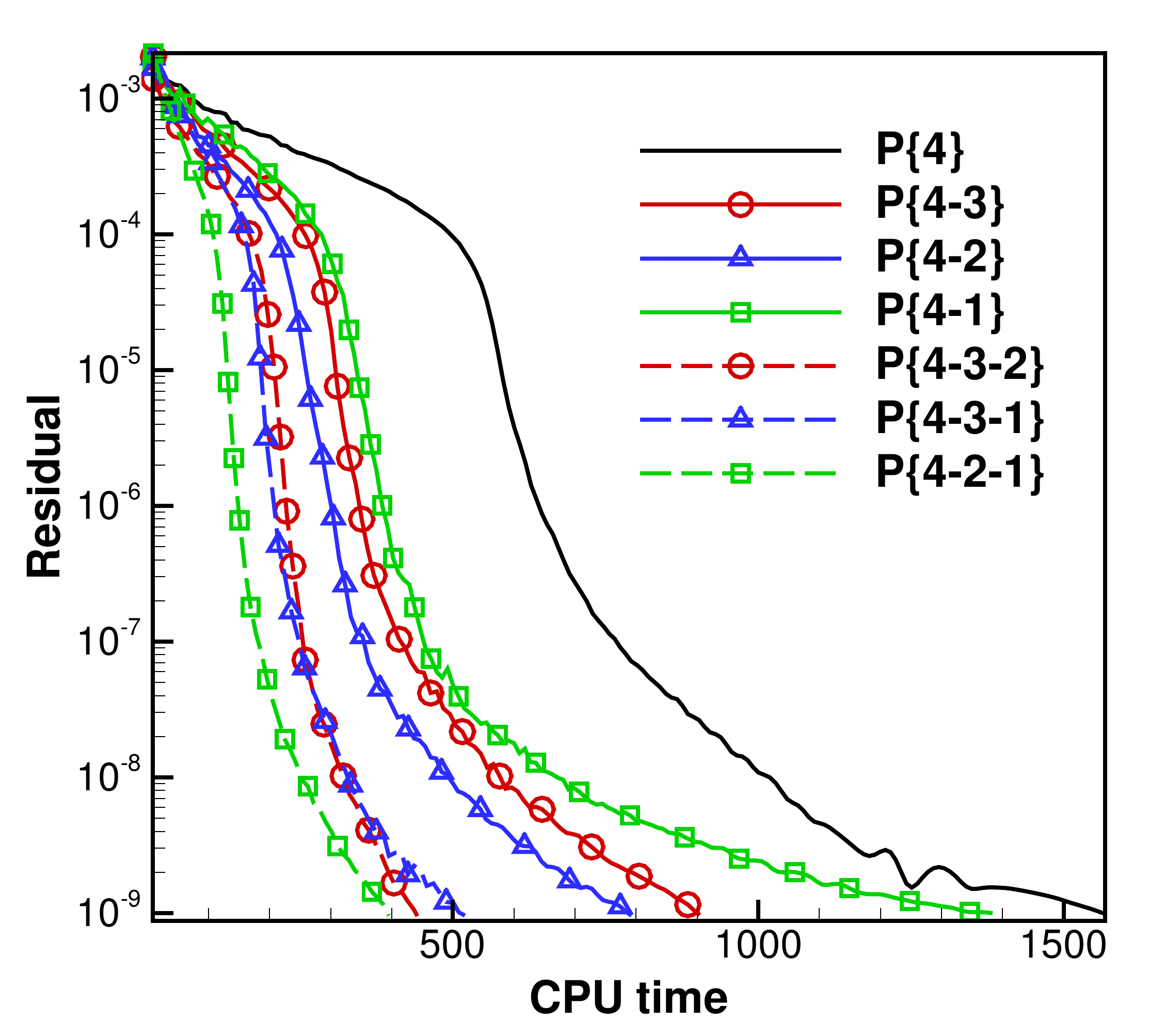}\\
		(a)& (b) \\	
	\end{tabular}
	\caption{Convergence histories of different $ P $-multigrid solvers for the $ P^4 $ FR discretization when solving the inviscid flow over a sphere at $\text{Ma}=0.001 $.}	
	\label{sphere_inv_p4}	
\end{figure}

In this subsection, we examine the convergence performance of $ P $-multigrid solvers with different hierarchies of polynomial degrees for 3D problems. We only consider the inviscid problem since it has been demonstrated in previous numerical experiments that the convergence performances of multigrid solvers for inviscid and convection-dominated viscous flows are similar.
The inviscid flow of $ \text{Ma}=0.001 $ over a sphere is studied here. To save computational cost, only a quarter of the sphere is considered. The mesh with 336 elements is shown in Figure~\ref{sphere_inviscid}. The curved surfaces are represented by $ P^3 $ elements.  The  normalized pressure and Mach number contours are given in Figure~\ref{sphere_inviscid}. The residual histories of the $ P^3 $ and $ P^4 $ FR discretization with different hierarchies of polynomial degrees are presented in Figure~\ref{sphere_inv_p3} and Figure~\ref{sphere_inv_p4}, respectively. In general, the performances of different $ P $-multigrid solver are consistent with previous findings.

\begin{figure}		
	\centering
	\begin{tabular}{cc}
		\includegraphics[width=6cm]{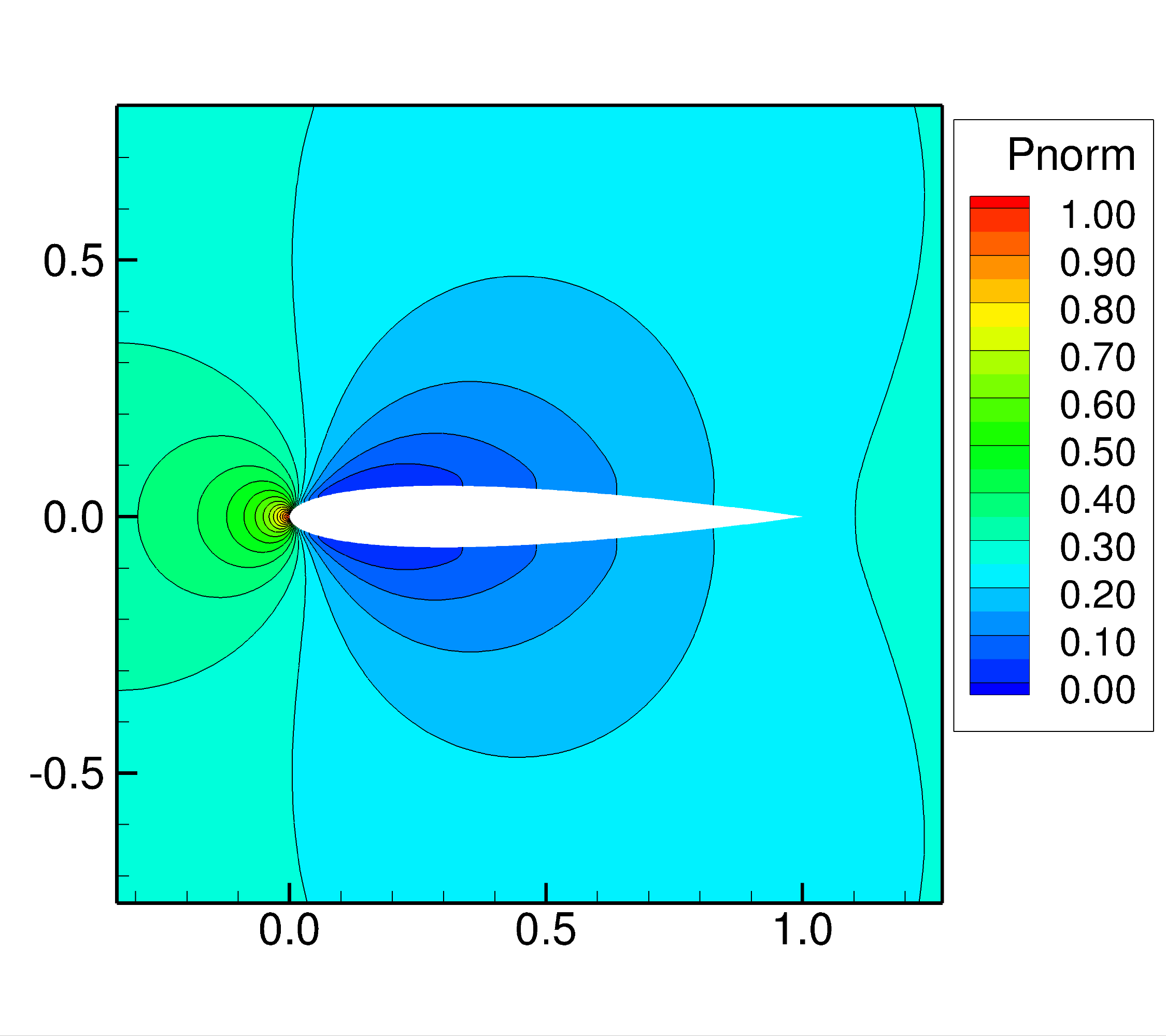} &
		\includegraphics[width=6cm]{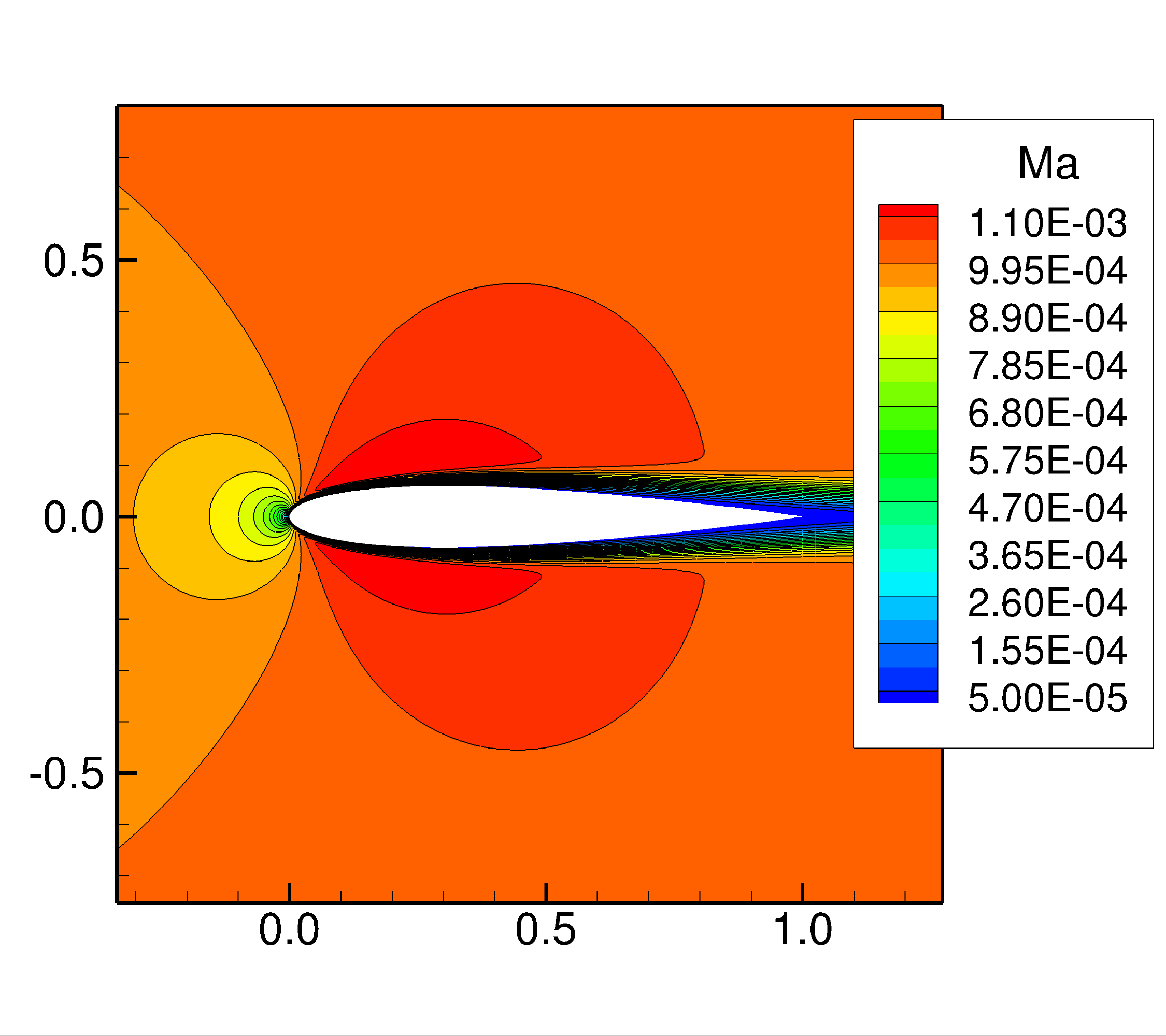}\\
		(a)& (b) \\	
	\end{tabular}
	\caption{(a) Normalized pressure field and (b) $ \text{Ma} $ number field of the viscous flow over a NACA0012 airfoil at $ \text{Ma}=0.001 $ and $ \text{Re}=5000 $.}	
	\label{naca_vis_contour}	
\end{figure}

\subsection{Under-resolved simulation of the transitional flows over an SD7003 wing}
\begin{figure}		
	\centering
	\begin{tabular}{cc}
		\includegraphics[width=6cm]{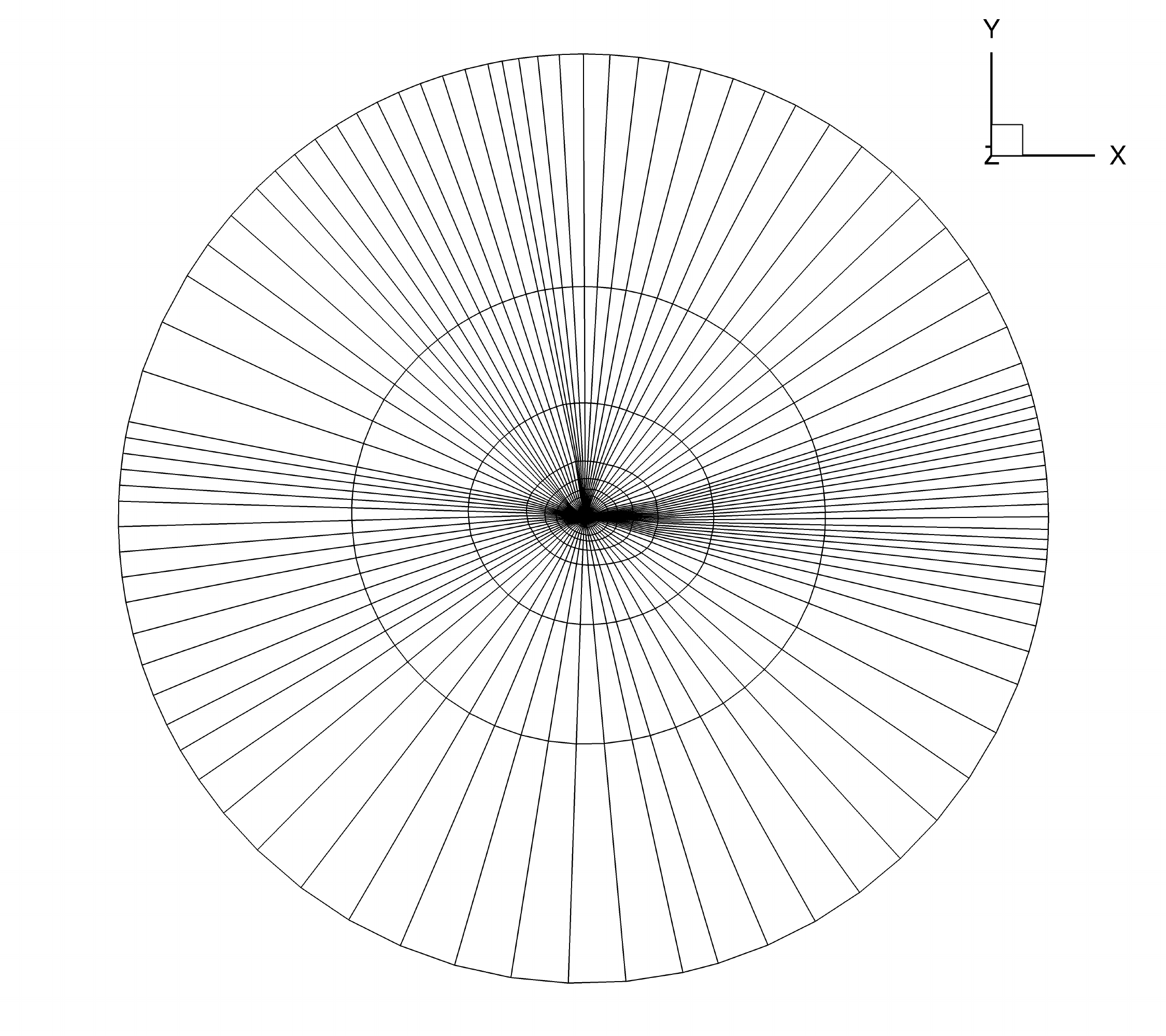} &
		\includegraphics[width=6cm]{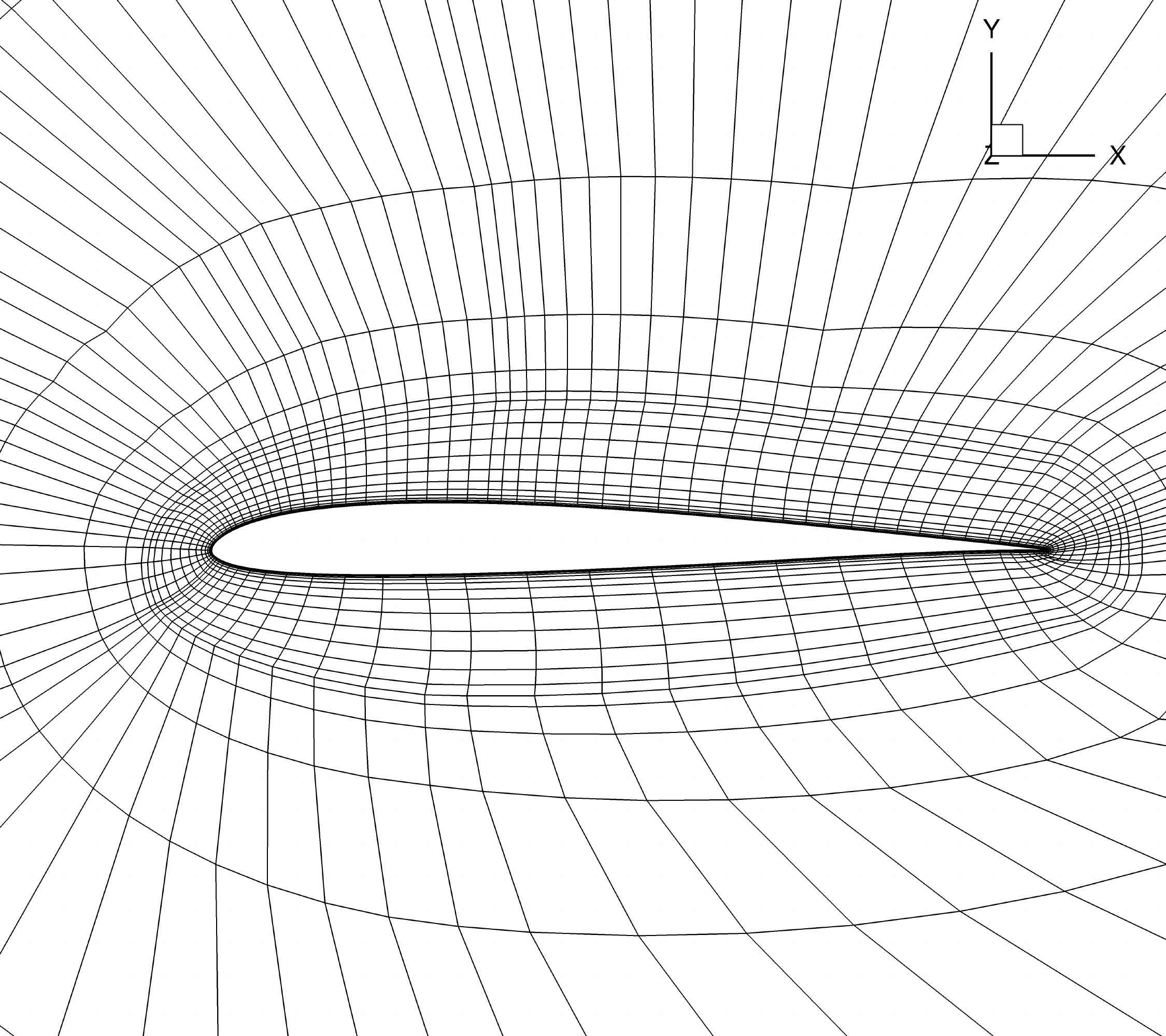}\\
		(a)& (b) \\	
	\end{tabular}
    
	\caption{Meshes around an SD7003 wing for under-resolved transitional flow simulation. (a) A global view and (b) a close-up view near the wing.}	
	\label{sd7003_mesh}	
\end{figure}

As a last step, we apply the $ P $-multigrid solver to simulate the transitional flows over an SD7003 wing. The Reynolds number of the inflow based on the chord length $ C $ of the wing is $ 60000 $. The angle of attack of the inflow is $ 8^\circ $. Two Mach numbers are studied, namely $ \text{Ma}=0.1 $ and $ \text{Ma}=0.01 $. The mesh employed for this study is illustrated in Figure~\ref{sd7003_mesh}. There are 26320 hexahedral  elements in total. The height of the first layer near the wing is $ 0.0003C $. Quadratic elements are used to represent the curved wall boundaries. The mesh is obtained by extruding the 2D mesh along the spanwise direction by $ 0.2C $ and 10 layers are allocated in the spanwise direction.

A three-level $ P $-multigrid solver, i.e., $ P\{3-2-1\} $, is used to solve the nonlinear systems. When $ \text{Ma}=0.1 $, $ I\{5-5-10\} $ is employed for the smoothing procedure at each level and when $ \text{Ma}=0.01 $, $ I\{10-10-16\} $ is used. The physical time step size is set as $ \Delta t = 0.002 $ for both simulations. Instead of providing $ CFL $ for the pseudo transient continuation, we set $ \Delta \tau_{init}  $ and $ \Delta \tau_{max} $ directly as $ \Delta \tau_{init} = 0.001 $ and $ \Delta \tau_{max} = 0.05 $ for both cases. When $ \text{Ma}=0.1 $, we update the element Jacobi smoother every 10 steps; when $ \text{Ma}=0.01 $, we update it every 20 steps. If $ \Delta \tau = \Delta \tau_{max} $, the smoother  will not be updated anymore. We require the relative residual of the pseudo transient continuation to drop three orders, i.e., $ tol_{pseudo} = 10^{-3} $, during each physical time step. 
The second order backward differentiation formula (BDF2) is used to perform simulations until $ t =3 $ to obtain the initial flow fields. ESDIRK2 is then employed to restart both simulations while the physical time is reset as  $ t=0 $. The instantaneous solutions in the time slot $ t \in [14,18] $ are used for statistics. The global cut-off parameter $ \kappa $ in Eq.~\eqref{eps_nonmove} is set as 1.0 for $ \text{Ma}=0.1 $ and $ 2.5 $ for $ \text{Ma}=0.01 $. A slightly larger $ \kappa $ for $ \text{Ma}=0.01 $ can accelerate the convergence speed.

\begin{figure}		
	\centering
	\includegraphics[width=10cm]{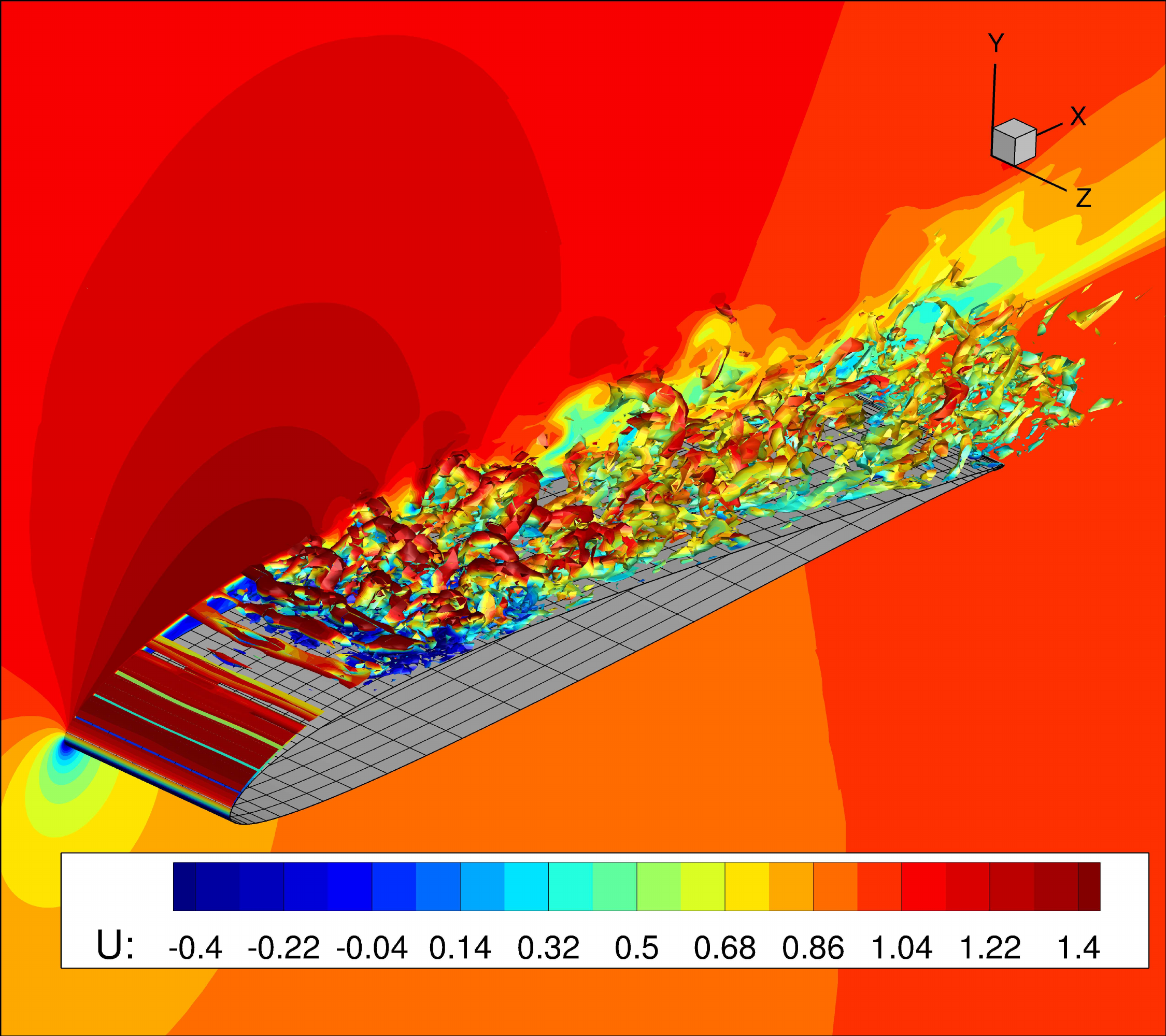} 	
	\caption{Iso-surfaces of the Q-criterion colored by the instantaneous streamwise velocity $ u $ at $ \text{Ma}=0.01 $. In this case, the value of Q is set as 500. }	
	\label{sd7003_Q}	
\end{figure}

\begin{figure}		
	\centering
	\begin{tabular}{cc}
		\includegraphics[width=7cm]{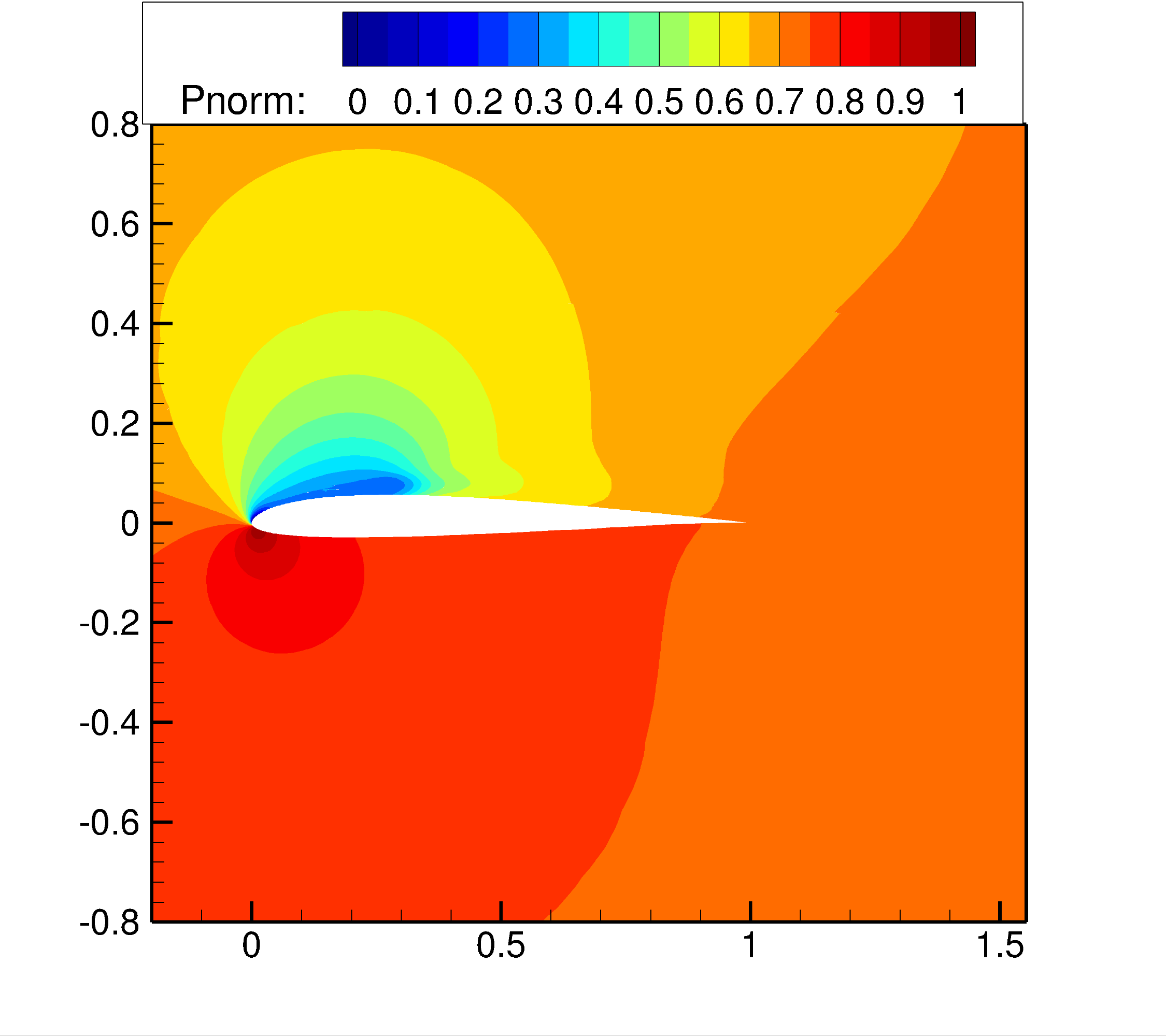} &
        \includegraphics[width=7cm]{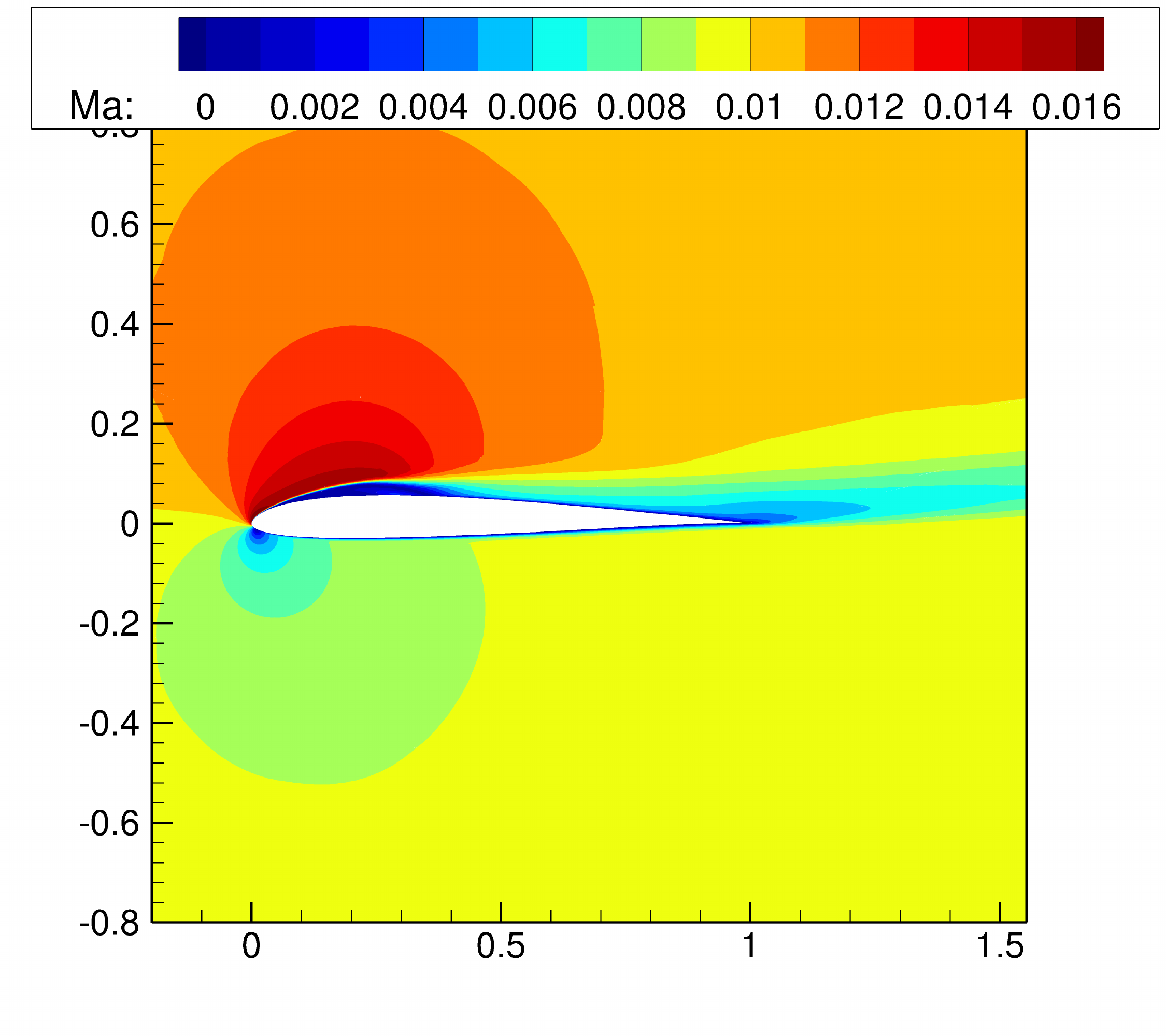}\\
		(a)& (b) \\	
	\end{tabular}
	
	\caption{Mean flow fields of the transitional flow over an SD7003 wing at $ \text{Ma} =0.01 $. (a) Contour of the normalized pressure $ p_{norm} $ and (b) contour of the Mach number.}	
	\label{sd7003_contour_ma0d01}	
\end{figure}

\begin{figure}		
	\centering
	\begin{tabular}{cc}
		\includegraphics[width=6cm]{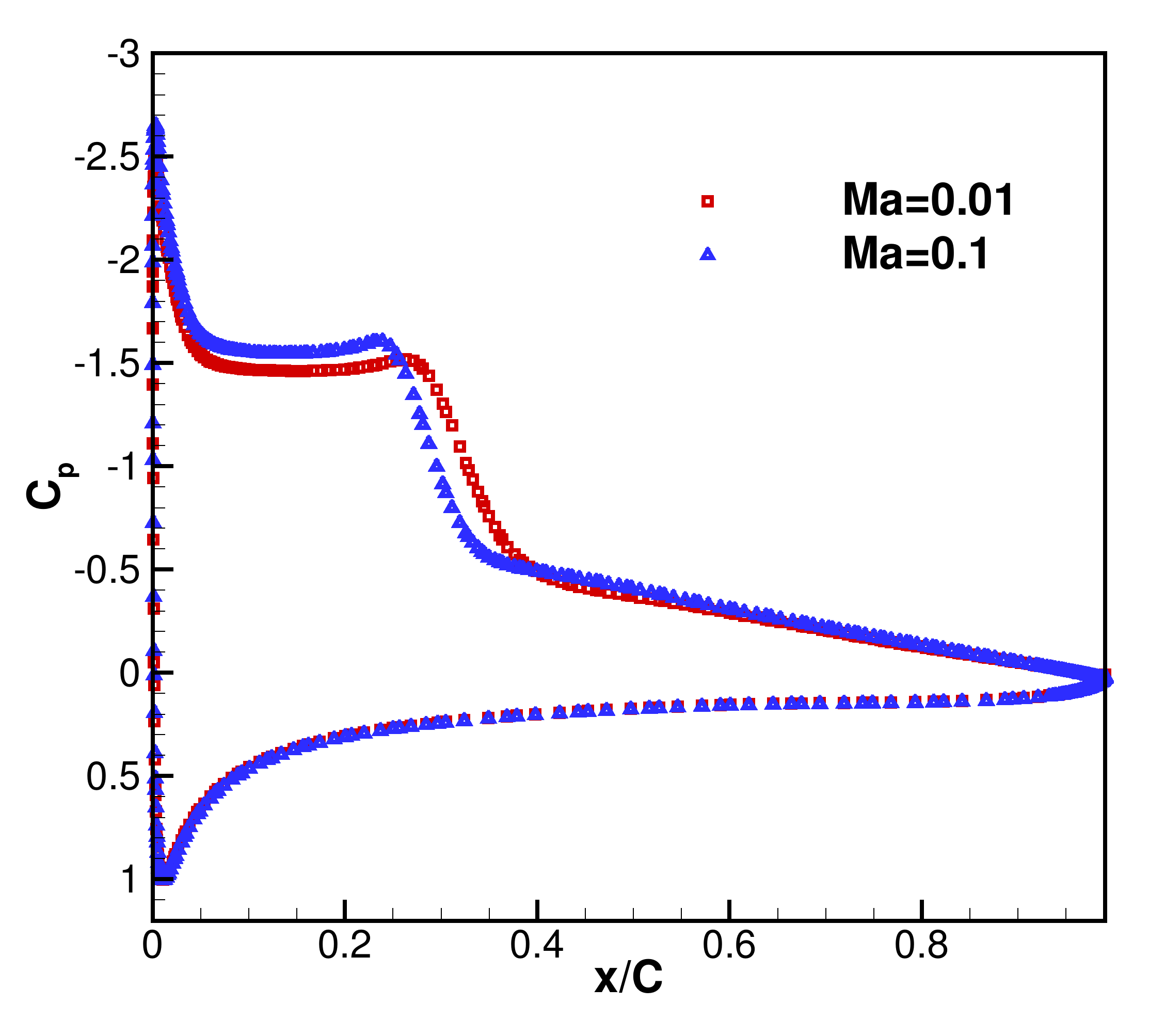} &
		\includegraphics[width=6cm]{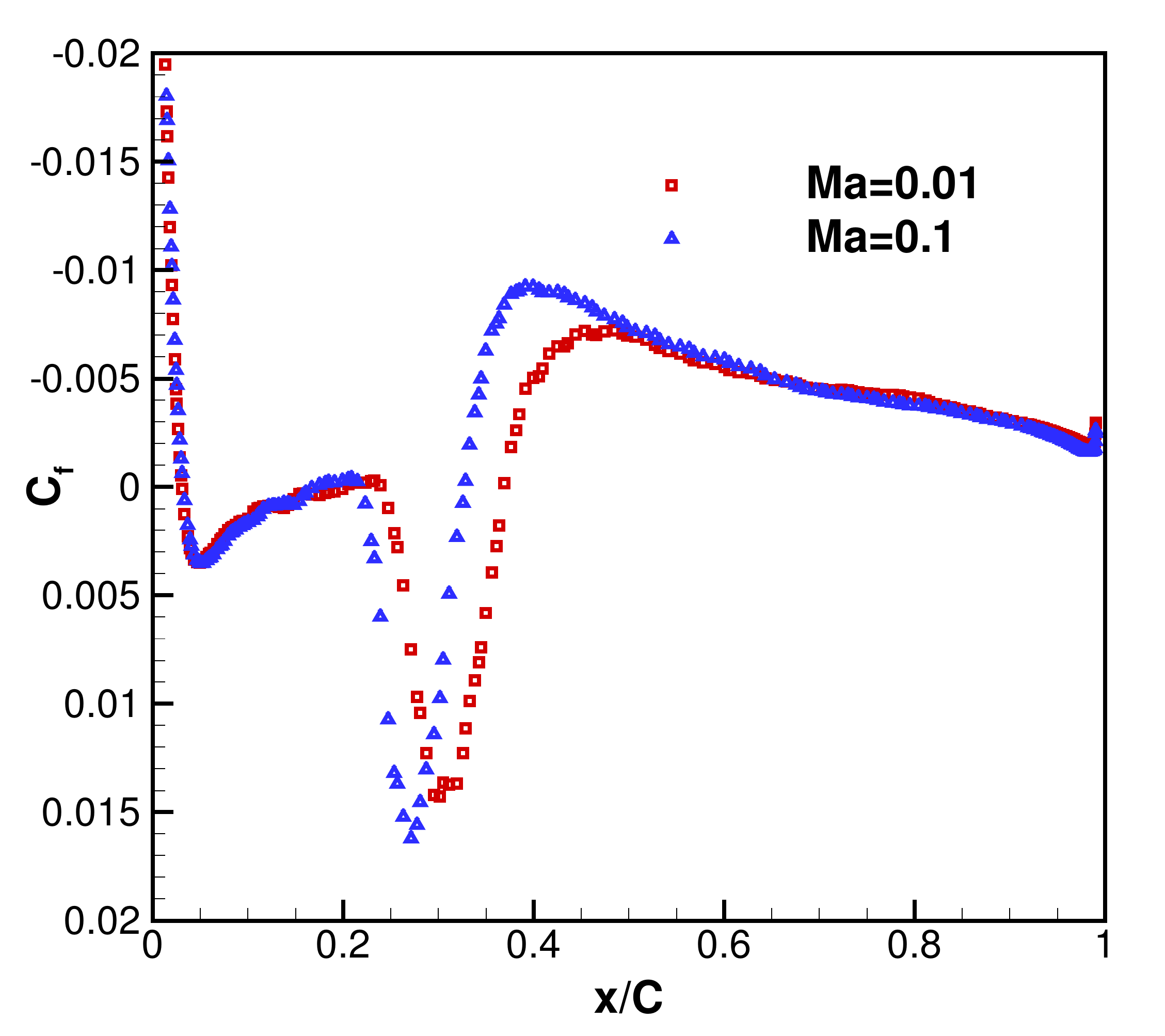}\\
		(a)& (b) \\	
	\end{tabular}
	
	\caption{(a) Time-averaged surface pressure coefficient $ C_p $ and (b) time-averaged surface friction coefficient $ C_f $ of the suction side.}	
	\label{sd7003_surface_coef}	
\end{figure}

An instance of the isosurface of the Q-criterion where $ Q=500 $ colored by the streamwise velocity $ u $ when $ \text{Ma} = 0.01 $ is presented in Figure~\ref{sd7003_Q}. The corresponding time-averaged fields of normalize pressure  and $ \text{Ma} $ are presented in Figure~\ref{sd7003_contour_ma0d01}. We note that the time-averaged flow fields when $ \text{Ma} = 0.1 $ are very similar to those when $ \text{Ma} = 0.01 $. Therefore, they are not presented here for brevity.
The time-averaged surface pressure coefficient $ C_p $ and friction coefficient $ C_f $ of the suction side are illustrated in Figure~\ref{sd7003_surface_coef}. The predicted $ C_l $, $ C_d $, separation point $ x_s $ and reattachment points $ x_r $ are documented in Table~\ref{sd7003_forces}. The lift predictions of our current work have a good agreement with the previous experimental and numerical results. All numerical studies over-predict the drag compared to the experiment by Selig et al.~\cite{selig1995summary}. A general trend in previous numerical studies is that when the Mach number becomes smaller, the lift generation will decrease. This trend has been observed in the current work. Few works can be found in literature using high-order methods to simulate this problem at $ \text{Ma} = 10^{-2} $.  At low speeds, the artificial compressibility method is usually employed with high-order methods~\cite{bassi2015linearly}. In our current work, when the $ \text{Ma} $ decreases from $ \text{Ma}=0.1 $ to $ \text{Ma}=0.01 $, the separation bubble becomes longer, and the lift production is reduced. It is not clear whether this is due to the dissipation introduced by larger $ \kappa $ or the weak compressibility of fluid at different Mach numbers. 
We note that in our previous work, it has been demonstrated that weak compressibility has a non-negligible effect on thrust generation of flapping wings at low Mach numbers~\cite{wang2019implicit}. More investigations remain to be conducted to unravel this open question.  

\begin{table}
	
	\centering
	\caption{A comparison of the predicted $ C_d $, $ C_l $, separation points $ x_s $ and reattachment point $ x_r $ of the transitional flow over an SD7003 wing between the current study and previous ones.} 
	\label{sd7003_forces}
	\small
	\begin{tabular}{rrrrrrr }		
		\hline
		\hline
		           & Condition        &$C_l $  & $ C_d$ &$ x_s/C $& $ x_r/C $&Method \\
		\hline
		Current &$ \text{Ma}=0.1 $ &0.920&0.048&0.032  &0.326&    $ P^3 $ FR\\
		Current&$ \text{Ma}=0.01 $&0.913 &0.053&0.030  &0.364& $ P^3 $ FR\\
		Vermeire et al.~\cite{vermeire2017utility} &$ \text{Ma}=0.2 $&0.941&0.049& 0.045&0.315&$ P^4 $FR\\
		Beck et al.~\cite{beck2014high} &$ \text{Ma}=0.1 $ &0.923& 0.045&0.027 &0.310& $ P^3 $ DG\\
		Beck et al.~\cite{beck2014high} & $ \text{Ma}=0.1 $&0.932&0.050&0.030&0.336& $ P^7 $ DG\\
		Galbriath \&Visbal~\cite{galbraith2008implicit}& $ \text{Ma}=0.1 $&0.91&0.043&0.04&0.28& $ O(h^6) $ FD \\
		Bassi et al.~\cite{bassi2015linearly}&Incompressible&0.962&0.042&0.027&0.268& $ P^3 $ DG\\
		Bassi et al.~\cite{bassi2015linearly} &Incompressible & 0.953  &0.045&0.027 &0.294&$ P^4 $ DG \\	
		Selig et al.~\cite{selig1995summary}&Experiment&0.92&0.029&&&\\
		\hline  
	\end{tabular}	
\end{table}

\begin{figure}		
	\centering
	\begin{tabular}{cc}
		\includegraphics[width=6cm]{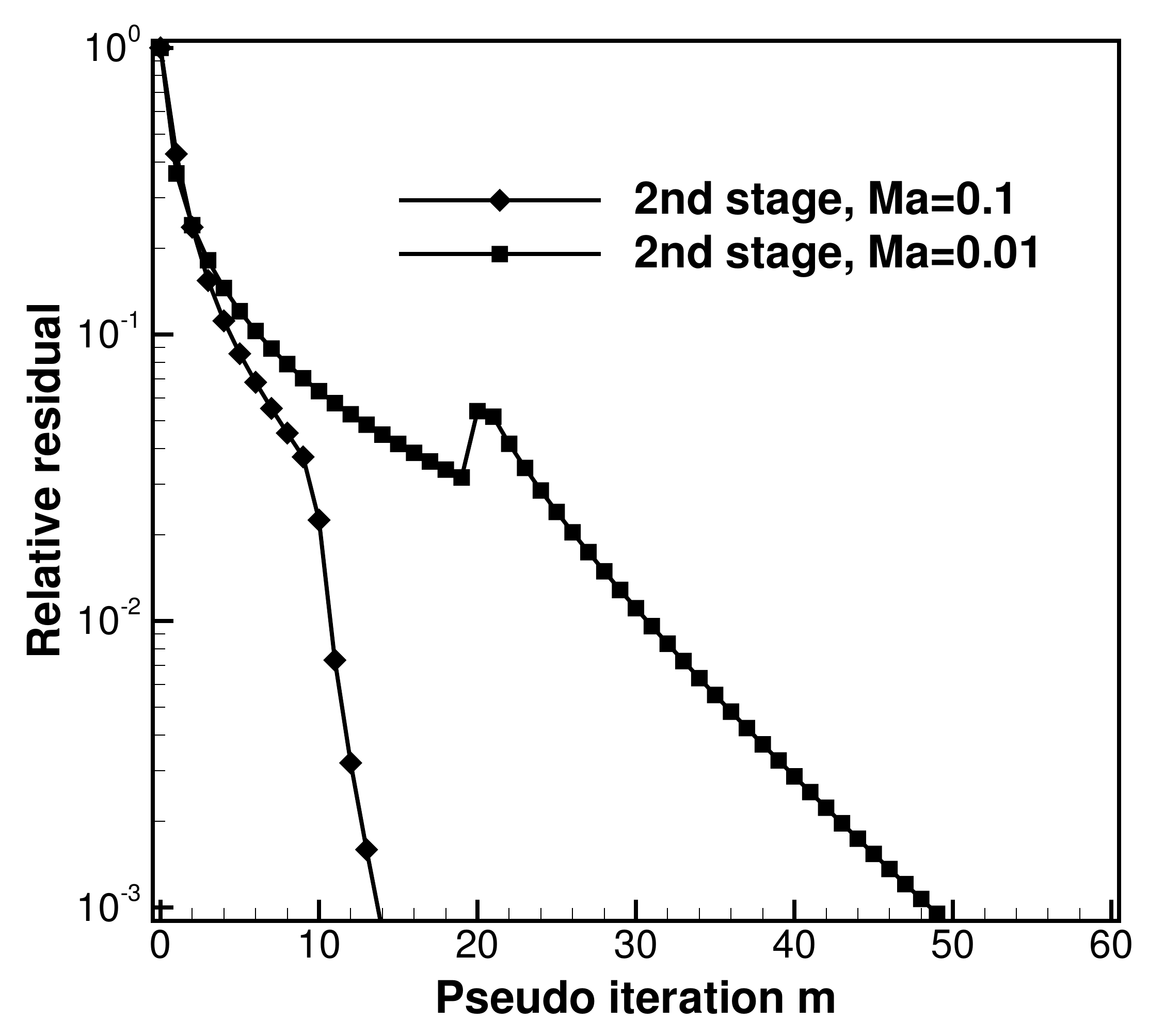} &
		\includegraphics[width=6cm]{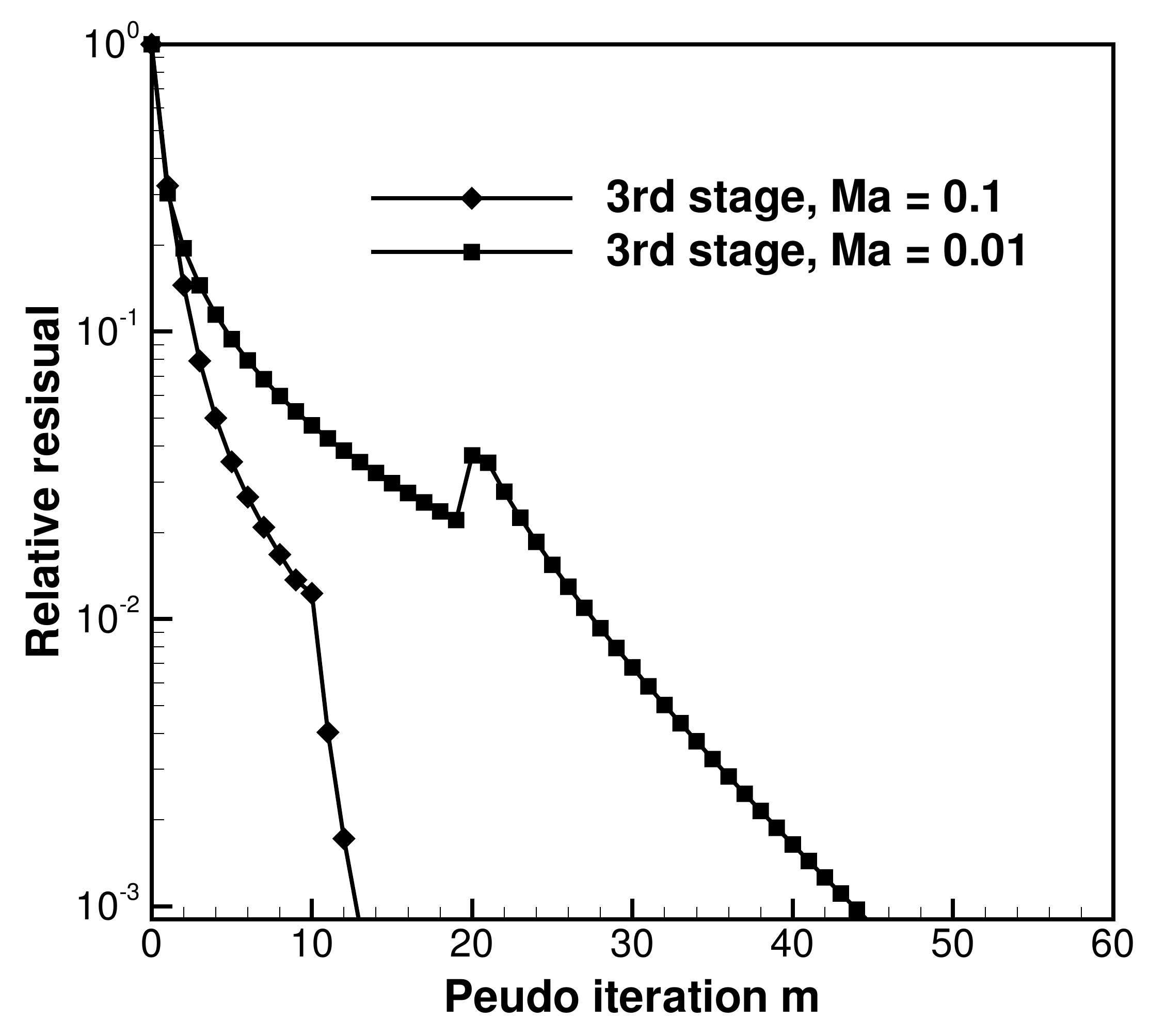}\\
		(a)& (b) \\	
	\end{tabular}
	
	\caption{Typical convergence histories of the relative residual for the pseudo transient continuation at the (a) second and (b) third stages of ESDIRK2 when simulating the transitional flow over an SD7003 wing.}	
	\label{sd7003_relres_convergence}	
\end{figure}

The stiffness of the compressible Navier--Stokes equations will significantly increase when the Mach number is reduced from $ 0.1 $ to $ 0.01 $. Therefore, it is expected that the $ P $-multigrid solver needs more iterations to converge. Instantaneous convergence histories of the relative residual for different $ \text{Ma} $ numbers when the flows are fully developed are shown in Figure~\ref{sd7003_relres_convergence}. When $ \text{Ma} = 0.1 $, the relative residual in the $2^{nd}$ stage and $3^{rd}$ stage of ESDIRK2 can be decreased by three orders of magnitude within 14 pseudo iterations; but when $ \text{Ma} = 0.01 $, 49 and 44 pseudo iterations are needed, respectively. Note that more iterations are used for the smoothing procedure at different levels for the $ P $-multigrid solver when $ \text{Ma} = 0.01 $ (recall that $ I\{5-5-10\} $ is used for $ \text{Ma}=0.1 $ and $ I\{10-10-16\} $ is used for $ \text{Ma}=0.01 $). This indicates that the computational cost of smoothing when $ \text{Ma}=0.01 $ is approximately as much as six times of that when $ \text{Ma}=0.1 $. Much effort is still needed to further improve the computational efficiency of the $ P $-multigrid solver for flow simulation at very low Mach numbers.

\section{Conclusions} \label{CFW}
We have developed a $ P $-multigrid solver to solve the nonlinear systems resulted from implicit high-order FR discretization of the locally preconditioned unsteady compressible Navier-Stokes equations at low Mach numbers. Specifically, high-order FR is employed for spatial discretization, and ESDIRK is employed for time integration. Local preconditioning is coupled with ESDIRK methods and is only enforced in the pseudo transient continuation procedure to preserve the accuracy of ESDIRK methods. High-order spatiotemporal accuracy is preserved for numerical simulation of low-Mach-number flows.

Through various numerical experiments, we found that for a two-level $ P $-multigrid solver, if the solver has a polynomial hierarchy close to $ \{P_0-P_0/2-P_0\} $, it would most likely have the best convergence performance. If the difference of the polynomial degrees between these two levels are excessively large, the correction from the coarser level may not only deteriorate the convergence speed, but also introduce new errors to the solution at the finer level. This can possibly lead to failure of convergence. Similarly, for a three-level $ P $-multigrid solver, a polynomial hierarchy close to $ \{P_0-P_0/2-P_0/4-P_0/2-P_0\} $ is suggested.

We have demonstrated the capability of the implicit high-order FR methods with $ P $-multigrid acceleration on conducting under-resolved turbulence simulation at moderate Reynolds numbers and low Mach numbers. Numerical results have a reasonable agreement with those from previous studies. We note that simulating massively turbulent flows at very low Mach numbers is challenging even with local preconditioning and $ P $-multigrid acceleration. To further accelerate turbulent flow simulation at very low Mach numbers, the Newton-Krylov methods with the $ P $-multigrid  preconditioner can be among promising candidates. This is our future work.

\section*{Acknowledgment}
The authors gratefully acknowledge the support of the Office of Naval Research through
the award N00014-16-1-2735, and the faculty startup support from the department of
mechanical engineering at the University of Maryland, Baltimore County (UMBC). The hardware used in the computational studies is part of the
UMBC High Performance Computing Facility (HPCF).
The facility is supported by the U.S. National Science Foundation
through the MRI program
(grant nos.~CNS-0821258, CNS-1228778, and OAC-1726023)
and the SCREMS program (grant no.~DMS-0821311),
with additional substantial support from UMBC.
\clearpage

\bibliographystyle{ieeetr}
\bibliography{p-multigrid}	
\end{document}